\documentclass{aa}  

\usepackage{graphicx}
\usepackage{amsmath}
\usepackage{txfonts}
\usepackage{hyperref}
\hypersetup{
    colorlinks,
    linkcolor={blue},
    citecolor={blue},
    urlcolor={blue}
}

\newcommand{\dr}{\delta_{\mathrm{r}}}
\newcommand{\dw}{\delta_{\omega}}
\newcommand{\ngrain}{n_{\mathrm{g}}}
\newcommand{\smin}{s_{\mathrm{min}}}
\newcommand{\smax}{s_{\mathrm{max}}}

\begin{document} 
\title{Dust production in the debris disk around HR\,4796\,A\thanks{Based on observations made with ESO Telescopes at the Paranal Observatory under program ID 097.C-0523, 082.C-0218, and 089.C-0207.}}

\author{J. Olofsson\inst{\ref{inst:IFA},\ref{inst:MPIA},\ref{inst:NPF}}
\and
J. Milli\inst{\ref{inst:ESO}}
\and
P. Th\'ebault\inst{\ref{inst:LESIA}}
\and
Q. Kral\inst{\ref{inst:LESIA}}
\and
F. M\'enard\inst{\ref{inst:IPAG}}
\and
M. Janson\inst{\ref{inst:Stockholm}}
\and
J.-C. Augereau\inst{\ref{inst:IPAG}}
\and
A. Bayo\inst{\ref{inst:IFA}, \ref{inst:NPF}}
\and
J. C. Beam\'in\inst{\ref{inst:AutoChile}, \ref{inst:NPF}}
\and
Th. Henning\inst{\ref{inst:MPIA}}
\and
D. Iglesias\inst{\ref{inst:IFA}, \ref{inst:NPF}}
\and
G. M. Kennedy\inst{\ref{inst:Warwick}}
\and
M. Montesinos\inst{\ref{inst:IFA}, \ref{inst:NPF}, \ref{inst:CASSACA}}
\and
N. Pawellek\inst{\ref{inst:MPIA}}
\and
M. R. Schreiber\inst{\ref{inst:IFA}, \ref{inst:NPF}}
\and
C. Zamora\inst{\ref{inst:IFA}, \ref{inst:NPF}}
\and
M. Carbillet\inst{\ref{inst:OCA}}
\and
P. Feautrier\inst{\ref{inst:IPAG}}
\and
T. Fusco\inst{\ref{inst:ONERA},\ref{inst:LAM}} 
\and
F. Madec\inst{\ref{inst:LAM}}
\and
P. Rabou\inst{\ref{inst:IPAG}}
\and
A. Sevin\inst{\ref{inst:LESIA}}
\and
J. Szul\'agyi\inst{\ref{inst:Zurich}}
\and
A. Zurlo\inst{\ref{inst:UDP}, \ref{inst:UDPinge}, \ref{inst:LAM}}
}

\institute{
Instituto de F\'isica y Astronom\'ia, Facultad de Ciencias, Universidad de Valpara\'iso, Av. Gran Breta\~na 1111, Playa Ancha, Valpara\'iso, Chile\\\email{johan.olofsson@uv.cl}\label{inst:IFA}
\and
Max Planck Institut f\"ur Astronomie, K\"onigstuhl 17, 69117 Heidelberg, Germany\label{inst:MPIA}
\and
N\'ucleo Milenio Formaci\'on Planetaria - NPF, Universidad de Valpara\'iso, Av. Gran Breta\~na 1111, Valpara\'iso, Chile\label{inst:NPF}
\and
European Southern Observatory (ESO), Alonso de C\'ordova 3107, Vitacura, Casilla 19001, Santiago, Chile\label{inst:ESO}
\and
LESIA-Observatoire de Paris, UPMC Univ. Paris 06, Univ. Paris-Diderot, 92195, Meudon, France\label{inst:LESIA}
\and
Univ. Grenoble Alpes, CNRS, IPAG, F-38000 Grenoble, France\label{inst:IPAG}
\and
Department of Astronomy, Stockholm University, AlbaNova University Center, 106 91, Stockholm, Sweden\label{inst:Stockholm}
\and
Nucleo de Astroqu\'imica y Astrof\'isica, Instituto de Ciencias Qu\'imicas Aplicadas, Facultad de Ingenier\'ia, Universidad Aut\'onoma de Chile, Av. Pedro de Valdivia 425, 7500912 Santiago, Chile\label{inst:AutoChile}
\and
Department of Physics, University of Warwick, Gibbet Hill Road, Coventry CV4 7AL, UK\label{inst:Warwick}
\and
Chinese Academy of Sciences South America Center for Astronomy, National Astronomical Observatories, CAS, Beijing 100012, China\label{inst:CASSACA}
\and
Universit\'e C\^ote d'Azur, OCA, CNRS, Lagrange, France\label{inst:OCA}
\and
DOTA, ONERA, Universit\'e Paris Saclay, F-91123, Palaiseau France\label{inst:ONERA}
\and
Aix Marseille Universit\'e, CNRS, LAM - Laboratoire d'Astrophysique de Marseille, UMR 7326, 13388, Marseille, France\label{inst:LAM}
\and
Center for Theoretical Astrophysics and Cosmology, Institute for Computational Science, University of Z\"urich, Winterthurerstrasse 190, CH-8057 Z\"urich, Switzerland\label{inst:Zurich}
\and
N\'ucleo de Astronom\'ia, Facultad de Ingenier\'ia y Ciencias, Universidad Diego Portales, Av. Ejercito 441, Santiago, Chile\label{inst:UDP}
\and
Escuela de Ingenier\'ia Industrial, Facultad de Ingenier\'ia y Ciencias, Universidad Diego Portales, Av. Ejercito 441, Santiago, Chile \label{inst:UDPinge}
}

\abstract
{Debris disks are the natural by-products of the planet formation process. Scattered or polarized light observations are mostly sensitive to small dust grains that are released from the grinding down of bigger planetesimals.}
{High angular resolution observations at optical wavelengths can provide key constraints on the radial and azimuthal distribution of the small dust grains. These constraints can help us better understand where most of the dust grains are released upon collisions.}
{We present SPHERE/ZIMPOL observations of the debris disk around HR\,4796\,A, and model the radial profiles along several azimuthal angles of the disk with a code that accounts for the effect of stellar radiation pressure. This enables us to derive an appropriate description for the radial and azimuthal distribution of the small dust grains.}
{Even though we only model the radial profiles along (or close to) the semi-major axis of the disk, our
best-fit model is not only in good agreement with our observations but also with previously published datasets (from near-IR to sub-mm wavelengths). We find that the reference radius is located at $76.4\pm0.4$\,au, and the disk has an eccentricity of $0.076_{-0.010}^{+0.016}$, with the pericenter located on the front side of the disk (north of the star). We find that small dust grains must be preferentially released near the pericenter to explain the observed brightness asymmetry.}
{Even though parent bodies spend more time near the apocenter, the brightness asymmetry implies that collisions happen more frequently near the pericenter of the disk. Our model can successfully reproduce the shape of the outer edge of the disk, without having to invoke an outer planet shepherding the debris disk. With a simple treatment of the effect of the radiation pressure, we conclude that the parent planetesimals are located in a narrow ring of about $3.6$\,au in width.
}


\maketitle
%

\section{Introduction}

Debris disks are the leftovers of stellar (and planetary) formation process(es). As the primordial gas-rich, massive circumstellar disk evolves, the small dust grains will grow and form planetesimals, which may become the building blocks of future planets. With a half-life time of a few million years (\citealp{Hernandez2007}), this primordial disk will rapidly transition towards its debris disk phase, losing the vast majority of its gaseous content. In this debris disk phase, the planetesimals leftover from the planet formation process collisionally erode to produce the small dust grains that are observed.

Recent advances in high contrast imaging instruments such as the \textit{Gemini Planet Imager} (GPI, \citealp{Perrin2015}) or the VLT/\textit{Spectro-Polarimetric High-contrast Exoplanet REsearch} (SPHERE, \citealp{Beuzit2019}) provide us with new avenues to investigate debris disks at high angular resolution. For instance, \citet{Olofsson2016} and \citet{Milli2017} studied the scattering phase function over a wide range of scattering angles for HD\,61005 and HR\,4796\,A, respectively, using the SPHERE instrument. Such studies help better constrain the nature of the small dust grains in young debris disks. \citet{Lee2016} presented numerical simulations that can explain a variety of morphologies of debris disks. Their code, which can account for the effect of stellar radiation pressure, was used by \citet{Esposito2016} to model GPI observations of the debris disk around HD\,61005. Overall, thanks to the exquisite spatial resolution provided by this new generation of instruments, we are able to perform in-depth studies of young and bright debris disks, trying to constrain the collisional activity responsible for the production of small dust grains.

Since \citet{Jura1991} reported the detection of mid- and far-infrared (IR) excess emission with the \textit{InfraRed Astronomy Satellite}, the debris disk around \object{HR\,4796\,A} has been intensely studied, from the ground and from space, with ever increasing image quality. HR\,4796\,A is young ($8\pm2$\,Myr, \citealp{Stauffer1995}) and nearby ($71.9 \pm 0.7$\,pc, \citealp{Gaia2018}). The A-type star, hosting one of the brightest debris disks (fractional luminosity of about $5 \times 10^{-3}$, \citealp{Moor2006}) is in a visual binary system (see Section\,\ref{sec:HR4796B}), with the secondary being an M-type star at a separation of $7.7\arcsec$ (\citealp{Jura1993}). Scattered light and thermal emission observations have revealed a narrow ring of dust at about 77\,au from the central A-type star (e.g., \citealp{Jayawardhana1998,Wyatt1999,Augereau1999,Wahhaj2005,Schneider2009,Thalmann2011,Moerchen2011,Lagrange2012,Rodigas2015,Milli2015,Perrin2015,Milli2017,Schneider2018,Kennedy2018,Milli2019}). Several studies have reported that the disk displays a brightness asymmetry along its major axis, which is most likely related to the non-zero eccentricity of the disk. It has been postulated that planets could be shepherding the parent planetesimals, inducing the observed azimuthal asymmetries via secular interactions (e.g., \citealp{Wyatt1999}). In this paper, we present SPHERE/ZIMPOL observations of HR\,4796\,A and we investigate the collisional activity and the production of small dust grains by constraining the azimuthal and radial distribution of the dust.

\section{Observations and data reduction}\label{sec:obs_data_red}

The data were obtained with SPHERE/ZIMPOL (\citealp{Schmid2018}) in its polarimetric mode P2 corresponding to field stabilized observations. They are part of the SPHERE Guaranteed Time Observations\footnote{ESO program 097.C-0523(A)}. The target HR\,4796\,A was observed on the night of 2016-5-24 during a sequence alternating between two short unsaturated polarimetric cycles in Fast Polarimetry (called Fast) and two deeper saturated polarimetric cycles in Slow Polarimetry (called Slow). We repeated three times the pattern Fast Slow Fast with the derotator position angle set to $0^\circ$, $30^\circ$ and $60^\circ$, respectively. Two more cycles were actually observed, with position angles of $120^{\circ}$ and $150^{\circ}$, but not used in this paper. For those last two cycles, columns of deeply saturated pixels fall right along the semi-major axis of the disk, and excluding them leads to a cleaner final reduction (but noisier in the regions close to the semi-minor axis). Given that the purpose of this paper is to study the radial profiles along the semi-major axes of the disk, we opted to exclude those additional cycles. This strategy enables us to get unsaturated frames bracketing the deep saturated exposures in order to calibrate the photometry, and the different derotator position angles aim at introducing an additional diversity parameter to smooth out any residual pattern. In this paper, we focus only on the polarized image of the disk without any absolute flux calibration, therefore only the deep unsaturated Slow Polarimetry images are used for the image discussed later. The absolute polarized flux of the disk and polarized fraction are treated in a separate paper (\citealp{Milli2019}). The extreme adaptive optics (SAXO; \citealp{Fusco2006}) yielded a point-spread function (PSF) with a full width at half maximum of $30-40$\,mas, larger than the expected diffraction limit because of the low-wind effect (see also \citealp{Milli2019}).

The images were reduced with custom Python routines. No calibration frames (such as dark, bias or flat field) are applied to avoid introducing additional sources of noise that might degrade the polarized sensitivity. The ZIMPOL instrument is indeed designed to beat atmospheric and instrumental speckles by using the concept of polarimetric differential imaging with a masked CCD synchronized with a ferroelectric liquid crystal modulated at a rate of 27\,Hz in Slow Polarimetry (\citealp{Schmid2012}). This unique design allows the signal coming from the two orthogonal polarization directions to be captured by the exact same pixels, self-calibrating any differential and pixel-to-pixel effect after applying the polarimetric subtraction. A half-wave plate (HWP) is introduced very early in the optical train to calibrate the instrumental polarization. Each polarimetric cycle includes four positions of the HWP: $0^\circ$, $22.5^\circ$, $45^\circ$ and $67.5^\circ$. We applied the double difference technique to obtain the two Stokes parameter $Q$ and $U$ out of these 4 HWP positions (\citealp{Avenhaus2014}). Additional instrumental polarization coming upstream from the HWP remains and are removed by subtracting from the Stokes $Q$ and $U$ a scaled version of the intensity image $I$, as described in \citet{Ginski2016} and \citet{Canovas2011}. We then constructed local Stokes vectors $Q_\phi$ (shown in Fig.\ref{fig:cuts}) and $U_\phi$, containing the astrophysical signal and an estimate of the noise, respectively (\citealp{Benisty2015,Olofsson2016}, see also \citealp{Canovas2015} for a discussion on the effect of optical depth).

\section{Grain-size dependent dust distribution}\label{sec:rp_code}

Scattered light observations in the near-infrared are sensitive to the small-end of the grain size distribution (e.g., \citealp{Mulders2013}). Therefore, to properly characterize the dust production and the collisional activity, one needs an appropriate prescription for the behavior of the $\mu$m-sized dust grains. Besides gravity, the dominant effect that can affect the orbital parameters of small dust grains is stellar radiation pressure (\citealp{Wyatt2005} showed that Poynting-Robertson drag does not play a significant role in massive debris disks, but see \citealp{Kennedy2015} for a discussion about detecting dust in the inner regions using nulling interferometry). Consequently, in this paper, we try to constrain the radial and azimuthal distribution of the dust, taking into account the effect of radiation pressure. 

\subsection{Description of the code}\label{sec:model}

The code used in this study is inspired by the work presented in \citet{Lee2016}. We start with an analytical description of the belt of planetesimal-sized parent bodies that produce the observed dust. The ring is defined by 6 parameters: a reference radius $r_0$, eccentricity $e$, position angle on the sky $\phi$ (positive from north to east), inclination $i$, argument of periapsis $\omega$, and width of the belt $\dr$. The radial distribution of the parent bodies follows a normal distribution centered at $r_0$ with a standard deviation $\dr$. All the dust grains, produced in a collisional cascade will originate from those parent bodies, whose sizes, numbers, or masses do not need to be explicitly defined in the code. However, the number of grains of different sizes released from those unseen parents bodies are defined by a size distribution between a minimum and a maximum size. Within this range of sizes, the grains are affected by radiation pressure from the star, and their dynamical evolution is size-dependent. One has to note that all parent bodies have a ``forced eccentricity'', as they all share the same $e$ and the same $\omega$.

As input parameters, we consider a grain size distribution, which initially follows the ``ideal'' prescription of \citet[][a differential power-law of the form d$n(s) \propto s^{-3.5}$d$s$, where $s$ is the grain size]{Dohnanyi1969}. The distribution is divided in $\ngrain$ intervals between the minimum and maximum grain sizes ($\smin$ and $\smax$, respectively). The number of grains in each interval is computed as Eq.\,2 of \citet{Dullemond2008}
\begin{equation}\label{eqn:ndens}
n(s) = \left( \frac{s}{\smin} \right) ^{p} \times s \times \Delta \mathrm{log}(s),
\end{equation}
where $\Delta \mathrm{log}(s)$ is the width of each bin in logarithmic space. Since the $\ngrain$ grain sizes are logarithmically spaced, each $\Delta \mathrm{log}(s)$ is the same, except the first and last ones which are half of that value (so that the grain size distribution is exactly between $\smin$ and $\smax$). For each grain size, we then compute the dimensionless $\beta$ ratio between radiation pressure and gravitational forces (\citealp{Burns1979}). For a given $s$, the value of $\beta$ depends on the dust properties (optical constants and density) and the stellar properties (mass and luminosity), and is evaluated as
\begin{equation}
    \beta(s) = \frac{3 L_\star}{16 \pi G c^2 M_\star} \times \frac{Q_{\mathrm{pr}}(s)}{\rho s},
\end{equation}
where $L_\star$ and $M_\star$ are the stellar mass and luminosity, $G$ the gravitational constant, $\rho$ the dust density. The radiation pressure efficiency $Q_\mathrm{pr}(s)$ is equal to $Q_\mathrm{ext}(s, \lambda) - g_\mathrm{sca}(s) \times Q_\mathrm{sca}(s, \lambda)$ averaged over the stellar spectrum, with $Q_\mathrm{ext}$ and $Q_\mathrm{sca}$ the extinction and scattering efficiencies (computed using the Mie theory), and $g_\mathrm{sca}$ the asymmetry parameter (the integral over $4\pi$ steradian of the phase function times the cosine of the scattering angle).

To decide where the collision releasing a dust grain takes place, we use a prior distribution on the mean anomaly. This ``collisional distribution'' can either be uniform or a normal distribution (centered either at the pericenter or the apocenter). The standard deviation when using the normal distribution is noted $\dw$. This implies that dust grains are not released uniformly in the disk, but that they can be released preferentially in a localized region of the disk, depending on the azimuth. The mean anomaly is then converted to the true anomaly $\nu$, by solving the Kepler equation. The effect of radiation pressure on the orbital parameters of a single dust grain is parametrized as in \citet{Wyatt1999,Wyatt2006,Lee2016}. Assuming that the parameters of the dust grain, upon its release, are $a$ (drawn from the normal distribution of width $\dr$ centered at $r_0$), $e$, $\beta$, and $\omega$, then its ``updated'' orbital parameters ($a_{\mathrm{n}}$, $e_{\mathrm{n}}$, and $\omega_{\mathrm{n}}$) are computed as
\begin{equation}\label{eqn:orbit}
\begin{aligned}
a_{\mathrm{n}} = \cfrac{a \times (1 - \beta)}{1 - 2\beta \cfrac{1 + e \mathrm{cos}(\nu)}{1 - e^2} }, \\
e_{\mathrm{n}} = \frac{1}{1 - \beta} \times \sqrt{e^2 + 2\beta e \mathrm{cos}(\nu) + \beta^2}, \\
\omega_{\mathrm{n}} = \omega + \mathrm{arctan}\left[\frac{\beta \mathrm{sin}(\nu)}{\beta\mathrm{cos}(\nu) + e}\right].
\end{aligned}
\end{equation}
For a given $\beta$, we make $3\,000$ realizations of the prior collisional distribution. For each realization (giving a set of orbital parameters $a_{\mathrm{n}}$, $e_{\mathrm{n}}$, and $\omega_{\mathrm{n}}$), we check if the updated eccentricity $e_{\mathrm{n}}$ is larger or equal to zero and strictly smaller than unity to avoid hyperbolic orbits. Similarly to \citet{Lohne2017}, the blow-out size does therefore depend on where the grains are launched from (e.g., pericenter or apocenter). If the orbit is bound, the code then populates it with $500$ dust particles, uniformly distributed in mean anomaly. We finally then draw from a normal distribution of standard deviation $h/r = 0.04$ to account for the vertical dispersion on the disk. The opening angle is set to $0.04$ following \citet{Thebault2009}. The $(x, y, z)$ positions of each particle are registered, and we then find the pixel of the image that is closest to the $(x, y, z)$ values depending on the inclination and position angle of the disk (with the same pixel scale as the observations being modeled). We therefore produce number density maps for each value of $\beta$. 

The modeling strategy described above does not take into account an important effect discussed in \citet{Strubbe2006} and \citet{Thebault2008}, which is that small grains produced inside the belt on high eccentricity (bound) orbits will spend most of their time in the collision-free outer regions where they cannot be collisionally destroyed. This will significantly enhance their collisional lifetimes, and thus their number density as compared to what would be obtained by simply spreading out the number density obtained with Eq.\,\ref{eqn:ndens} over their whole orbit. To a first order, \citet{Strubbe2006} and \citet{Thebault2008} found that a correcting ``enhancement'' factor should be applied to the high-$\beta$ grain number density, which is roughly proportional to their total orbital period divided by the time spent within the birth ring. We take here the simplified expression for this correction factor given in \citet{Lee2016} and apply the following strategy. For each of $3\,000$ particles at a given $\beta$ that are not sent on hyperbolic orbits, we compute this correcting factor as $(1 - \beta)^{\alpha} / [1 - e^2 - 2 \beta \times (1 + e\mathrm{cos}(\nu))]^{\alpha}$, with $\alpha = 3/2$ ($e$ being the eccentricity of the parent belt, \citealp{Lee2016}), and $\nu$ the true anomaly at the moment of the collision. This correction should naturally produce a surface brightness profile of $r^{-3.5}$. However, such a profile is an asymptotic behavior that is reached relatively far away from the parent planetesimal belt, but not right after the birth ring (\citealp{Thebault2012}). When computing the number density maps for each $\beta$ values, the contribution of each particle is multiplied by this correction factor. One should also note that we do not take grain-grain collisions into account, which, as demonstrated in \citet{Lohne2017} can have an impact on the radial extent of the disk.

Once the $3\,000 \times 500$ particles have been launched, the scattering angle between the central star and the observer is computed for each pixel in the image. The code then computes one image per grain size bin by multiplying the number density of each pixel by $S_{12} \times \pi s^2 \times Q_{\mathrm{sca}}/(4 \pi r^2)$, where $r$ is the distance to the star, $Q_{\mathrm{sca}}$ the scattering efficiency, and $S_{12}$ the polarized phase function (which can be computed using the Mie theory, or other means). By using the $S_{12}$ element, we are effectively computing the $Q_{\phi}$ image directly, and not the $Q$ and $U$ images. The code can compute scattered light images replacing $S_{12}$ with the $S_{11}$ element of the M\"uller matrix. It can also compute thermal emission images by multiplying the number density map by $4 \pi s^2 Q_{\mathrm{abs}} \pi B_{\nu}(T_{\mathrm{dust}})$ (the dust temperature being evaluated by equating the energy received and emitted by a dust grain of size $s$ at the distance $r$ from the star, Eq.\,5 of \citealp{Wolf2005}). The final image is the collapse of all individual images, weighted by $n(s)$.

\subsection{Stellar parameters}

To derive the stellar parameters, we first gathered optical and near-IR photometry using the {\it VO SED Analyzer} (VOSA) tool\footnote{http://svo2.cab.inta-csic.es/theory/vosa50/} \citep{Bayo2008}. The stellar effective temperature and surface gravity were retrieved using the VO-DUNES discovery tool\footnote{http://sdc.cab.inta-csic.es/dunes/searchform.jsp} \citep{Eiroa2013}, which explores VIZIER catalogs. We find a value of $T_{\star} = 9\,700$\,K and a $\mathrm{log} (g) = 4.05$. Fitting a Kurucz model (\citealp{Castelli1997}) to the optical and near-IR photometry, we find a stellar luminosity of $L_{\star} = 25.75$\,L$_{\odot}$. The distance of $71.9$\,pc combined with the dilution factor used to scale the Kurucz model leads to a stellar radius of $R_{\star} = 1.79$\,R$_{\odot}$. The stellar mass is determined assuming that the star has finished contracting and we used the following relation between the stellar mass, radius and surface gravity,
\begin{equation}
\mathrm{log} (g) = 4.44 + \mathrm{log} \left( \frac{M_{\star}}{M_{\odot}} \right) - 2 \mathrm{log} \left( \frac{R_{\star}}{R_{\odot}} \right).
\end{equation}
We obtain a mass of $1.31$\,M$_{\odot}$.

\subsection{Disk parameters}

\citet{Milli2017} and \citet{Kennedy2018} presented comprehensive studies of the debris disk as seen in scattered light with VLT/SPHERE IRDIS and in thermal emission with ALMA, respectively. They constrain the main parameters of the disk, and because of our modeling strategy (see next sub-section), we used their results to set the inclination $i$ to $76.6^{\circ}$. \citet{Kennedy2018} reported a value of $76.6^{\circ}\pm0.2$, \citet{Milli2017} a value of $76.45^{\circ}\pm0.7$, \citet{Schneider2018} a value of $75.9^{\circ}\pm0.14$, \citet{Schneider2009} a value of $75.88^{\circ}\pm0.16$, \citet{Thalmann2011} a value of $76.7^{\circ}\pm0.5$. Overall, our choice for the inclination is consistent with previous studies, especially given that it was derived from high signal-to-noise observations at high angular resolution. For the position angle $\phi$, we used a maximum merit function as in \citet[][see \citealp{Olofsson2016} for details on how the elliptic mask is defined]{Thalmann2011}. We find that a value of $\phi = -151.6^{\circ}$ provides a best fit to the observations, but this value will be re-evaluated during the modeling of the observations. \citet{Milli2017} found $\phi = -152.9^{\circ}$, but one should note that we use a different convention for the position angle as the one used in \citet{Milli2017}. For $\phi = 0^{\circ}$ we assume that the major axis is along the north-south axis, with the near side of the disk being towards the east, hence the $180^{\circ}$ difference with the value of $+27.1^{\circ}$ reported in \citet{Milli2017}. One should also note that we do not constrain the direction in which the dust particles orbit around the star, and that the problem is symmetric.

\subsection{Modeling strategy}

\begin{figure}
\centering
\includegraphics[width=\hsize]{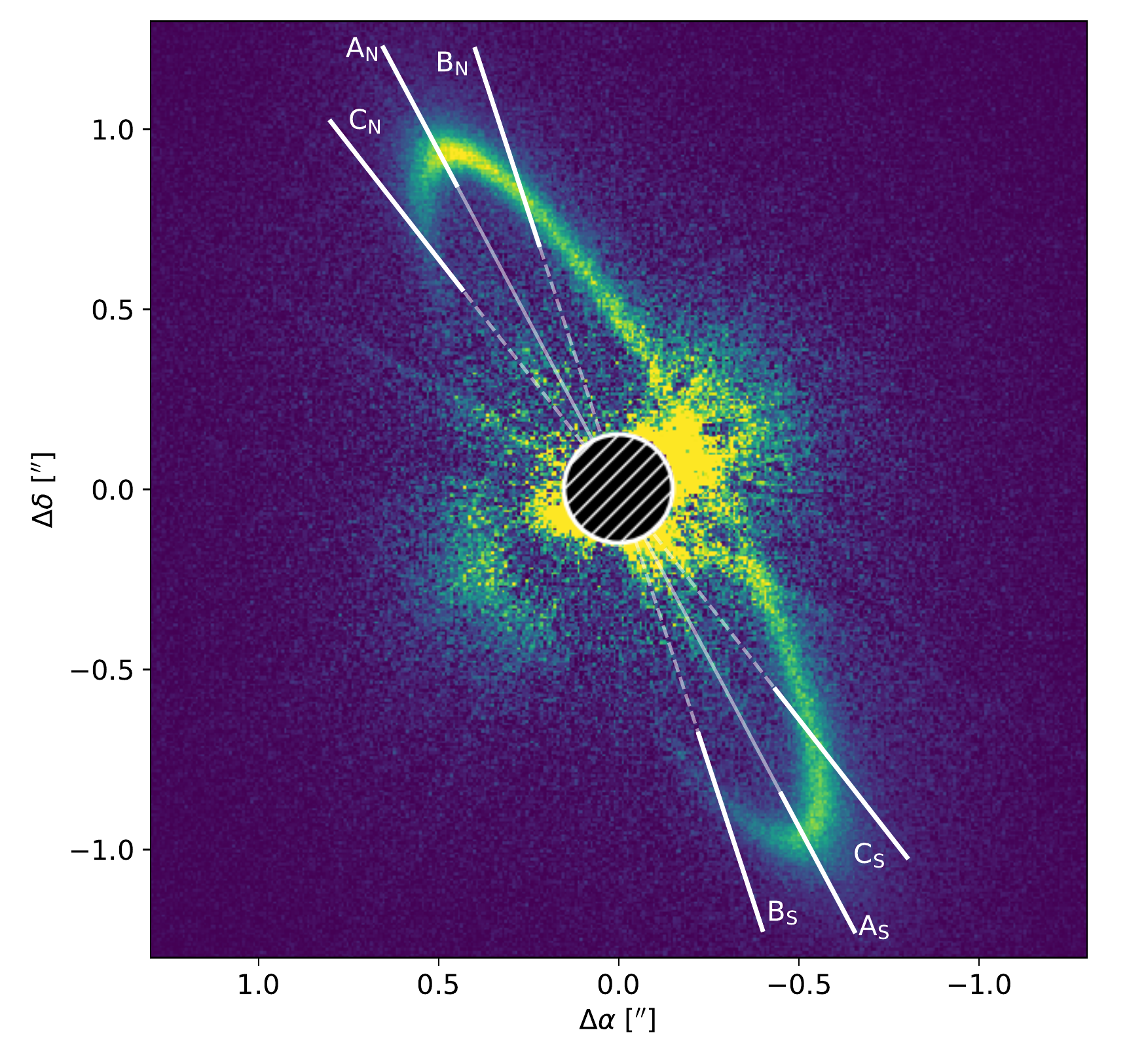}
    \caption{Reduced ZIMPOL image of HR\,4796\,A. The super-imposed lines A to C show the locations where we measure the radial profiles (the width of the lines does not correspond to the width of the slits used to measure the radial profiles).}
\label{fig:cuts}
\end{figure}

Choosing the adequate scattering theory to compute the full (polarization and scattering) phase function when modeling debris disks observations still remains a challenge. The Mie theory (commonly used in the literature as it is computationally fast) seems to be insufficient to reproduce most of the spatially resolved observations (\citealp{Lebreton2012,Rodigas2015,Olofsson2016,Milli2017}). Other alternatives exist, such as the discrete dipole approximation (\citealp{Purcell1973}), but can be time costly. Therefore, to alleviate this challenge, in this study we primarily focus on the radial profiles along the major axis of the disk. By doing so, and assuming that the disk is flat enough, we are probing the exact same scattering angles on both sides of the disk (close to $90^{\circ}$). Consequently, the exact value of the polarized phase function $S_{12}$ (integrated over all sizes) at this scattering angle does not matter when comparing the radial profiles along the north-east and south-west sides. The dependency of $S_{12}(90^{\circ})$ as a function of the grain size remains, but given that observations in the optical are mostly sensitive to the small dust grains, which are dominating in number, we consider this effect to be of second order. 

Nonetheless, modeling the radial cuts along the semi-major axis only does not allow us to really constrain the morphology of the disk. Preliminary results using solely radial cuts along the major axis led to clearly wrong results; the best fitting model would have a very large eccentricity ($e \sim 0.2-0.3$) and large reference radius ($r_0 \sim 100-120$\,au) so that the major axis of the model would not be the same as in the observations but the radial cuts would intercept the disk at other azimuthal angles. Therefore we considered two additional radial profiles at $\phi \pm 10^{\circ}$ from the major axis, to constrain the peak position of the disk, and obtain a more reliable determination of $r_0$, $e$, and $\omega$. We chose not to consider additional cuts closer to the semi-minor axis as the signal-to-noise degrades quite significantly. Moreover, our study focuses on determining the radial density profiles, which are best constrained close to the semi-major axis. Figure\,\ref{fig:cuts} shows the location of those radial cuts.

The observations are de-rotated to align each of the different axes with the vertical direction. The radial profiles are then computed as the average of the polarized flux (from the $Q_{\phi}$ image) over a vertical slit centered on the central pixel, with a width of $\pm\,5$\,pixels. For the uncertainties, we compute the standard deviation over the same vertical slit, from the $U_{\phi}$ image. For a given synthetic image, because of the finite number of particles that are used to generate the image, in some cases there can be some ``shot noise'', with a given pixel being much brighter than its neighbors. This may lead to artificial local minima when trying to find the best fit model. To circumvent this issue, the model image is first smoothed by performing a median clipping over $3 \times 3$ neighboring pixels (the central pixel, which value is being estimated, is not included when computing the median). It is then convolved by a 2D gaussian of standard deviation $1.22 \times \lambda/D$, where $D = 8$\,m the diameter of the telescope and $\lambda = 0.735$\,$\mu$m. We then proceed similarly to the observations to extract the radial profiles, and scale them by finding the scaling factor that minimizes the difference between the north-east and south-west sides simultaneously. For the profile along the major axis (A$_{\mathrm{N}}$ and A$_{\mathrm{S}}$ on Fig.\,\ref{fig:cuts}), the scaling factor is the same for both sides, as the polarized phase function should be the same. This allows us to test if the best fit model can successfully reproduce the brightness asymmetry between the north and south sides. For the radial cuts along B$_{\mathrm{N}}$, B$_{\mathrm{S}}$, C$_{\mathrm{N}}$, and C$_{\mathrm{S}}$, each scaling factor is determined independently as the polarized phase function is sampled at different angles (in principle, the phase function should be similar for the pairs C$_{\mathrm{N}}$-B$_{\mathrm{S}}$ and B$_{\mathrm{N}}$-C$_{\mathrm{S}}$ but we left the scaling factors unconstrained). Since the central region of the observations is contaminated by the instrumental PSF, we compute the goodness of fit between the ranges $\left[-1\farcs4, -0\farcs95\right]$ and $\left[0\farcs95, 1\farcs4\right]$ along the major axis, and $\left[-1\farcs3, -0\farcs7\right]$ and $\left[0\farcs7, 1\farcs3\right]$ for the other axis (these regions are highlighted by the wider white solid lines on Fig.\ref{fig:cuts}).

The free parameters of the model are the reference radius $r_0$, the standard deviation of the radial distribution $\dr$, the argument of periapsis $\omega$,  the standard deviation of the azimuthal collision probability at the pericenter $\dw$, the eccentricity $e$, and the position angle $\phi$. The location of the star is fixed and is not allowed to vary in our modeling approach. As mentioned before the inclination is not a free parameter and is set at $76.6^{\circ}$; the radial cuts are all close to the major axis of the disk, the worst direction to properly determine the inclination $i$. The grain size distribution exponent $p$ is set to $-3.5$. The grain size distribution is defined between $\smin = 6$\,$\mu$m (small enough that the grains in the first few bins are set on hyperbolic orbits) and $\smax = 1.3$\,mm. The value of $\smax$ is chosen so that the minimum value of $\beta$ is $5 \times 10^{-3}$ (optical constants of astrosilicates from \citealp{Draine2003}, with a density of $3.5$\,g.cm$^{-3}$) and that the grain size distribution is sampled over a significant range of sizes (we set $\ngrain = 200$). 

For the polarized phase function, we use the analytical Henyey-Greenstein expression, as
\begin{equation}
S_{12,\mathrm{HG}} = \frac{1 - \mathrm{cos}^2(\theta)}{1 + \mathrm{cos}^2(\theta)} \frac{1}{4\pi} \frac{1-g^2}{(1+g^2 - 2g\mathrm{cos}(\theta))^{3/2}},
\end{equation}
where $\theta$ is the scattering angle and $g$ the anisotropic scattering factor ($-1 \leq g \leq 1$). Similar approaches are commonly used for modeling polarimetric images of debris disks (e.g., \citealp{Engler2017}, who also included a term for the diffraction peak caused by grains larger than the wavelength). Nonetheless, in the preliminary versions of \citet{Milli2019} the authors found that the polarized phase function at scattering angles close to $90^{\circ}$, is overall well reproduced using $g \sim 0.3$\footnote{Since then, \citet{Milli2019} revised this value to $g = 0.4$, but with our modeling strategy, the exact value of $g$ is not really relevant. As mentioned before, we are fitting both sides of semi-major axis simultaneously (as they probe the same scattering angle the shape of the phase function does not matter), while for all the other radial cuts (B$_{\mathrm{N}}$, B$_{\mathrm{S}}$, C$_{\mathrm{N}}$, and C$_{\mathrm{S}}$), each profile is scaled up or down separately.} (the final best fit being obtained using a combination of two Henyey-Greenstein functions, to best match the brightness close to the semi-minor axis, which we do not try to reproduce here). We will therefore use this value of $g = 0.3$ throughout the paper, to alleviate the number of free parameters. Discussing the polarized phase function and the implications it has on the dust properties is out of the scope of this paper and we refer the reader to \citet{Milli2019}. The choice of the analytical form of the polarized phase function also alleviates some of the free parameters regarding the dust properties (such as the porosity as well, see \citealp{Arnold2019}). 
Indeed, now that the phase function is parametrized for all sizes, the absolute value of $\beta$ associated with each grain size matters less; what matters is the global shape of the $\beta(s)$ curve as it will determine the radial distribution of the dust grains, which we are modeling. With our approach we can therefore ignore most of the dust properties\footnote{The only remaining value that still depends on the choice of the scattering theory is the value of $Q_\mathrm{sca}$} and focus on the impact that the radiation pressure has on the spatial distribution of the small dust grains in the disk, regardless of their true sizes and properties.

To determine the best-fitting parameters and estimate their uncertainties, we then used an affine invariant ensemble sample Monte-Carlo Markov Chain (\texttt{emcee} package, \citealp{Foreman-Mackey2013}). The initial conditions are set to be close to the disk parameters reported in previous studies (for $r_0$, $e$, $\omega$, and $\phi$). We used $30$\,walkers and ran a short burn-in phase (length of $400$ models) and then run chains of $1\,000$ models for each walker. The uniform priors are also reported in Table\,\ref{tab:grid}. At the end, the mean acceptance fraction is of $0.27$ (indicative of convergence and stability, \citealp{Gelman1992}), and the maximum auto-correlation length among all the free parameters is $68$ steps.

\subsection{Results}\label{sec:results}

\begin{table}
\caption{Details for the modeling of the observations and best-fit results.}
\label{tab:grid}
\centering
\begin{tabular}{lcc}
\hline\hline
Parameters & Prior & Best-fit \\
\hline
$r_0$ [au]            & [$70$, $85$]     & $76.4_{-0.3}^{+0.4}$ \\
$\dr$ [au]            & [$1$, $5$]       & $3.6_{-0.2}^{+0.2}$ \\
$\omega$ [$^{\circ}$] & [$-270$, $-90$]  & $-254.3_{-1.6}^{+1.8}$ \\
$\dw$ [$^{\circ}$]    & [$20$, $120$]    & $63.9_{-11.3}^{+14.4}$ \\
$e$                   & [$0.0$, $0.2$]    & $0.076_{-0.010}^{+0.016}$ \\
$\phi$ [$^{\circ}$]   & [$-158$, $-148$] & $-152.1_{-0.1}^{+0.1}$ \\
\hline
\end{tabular}
\end{table}

\begin{figure}
\centering
\includegraphics[width=\hsize]{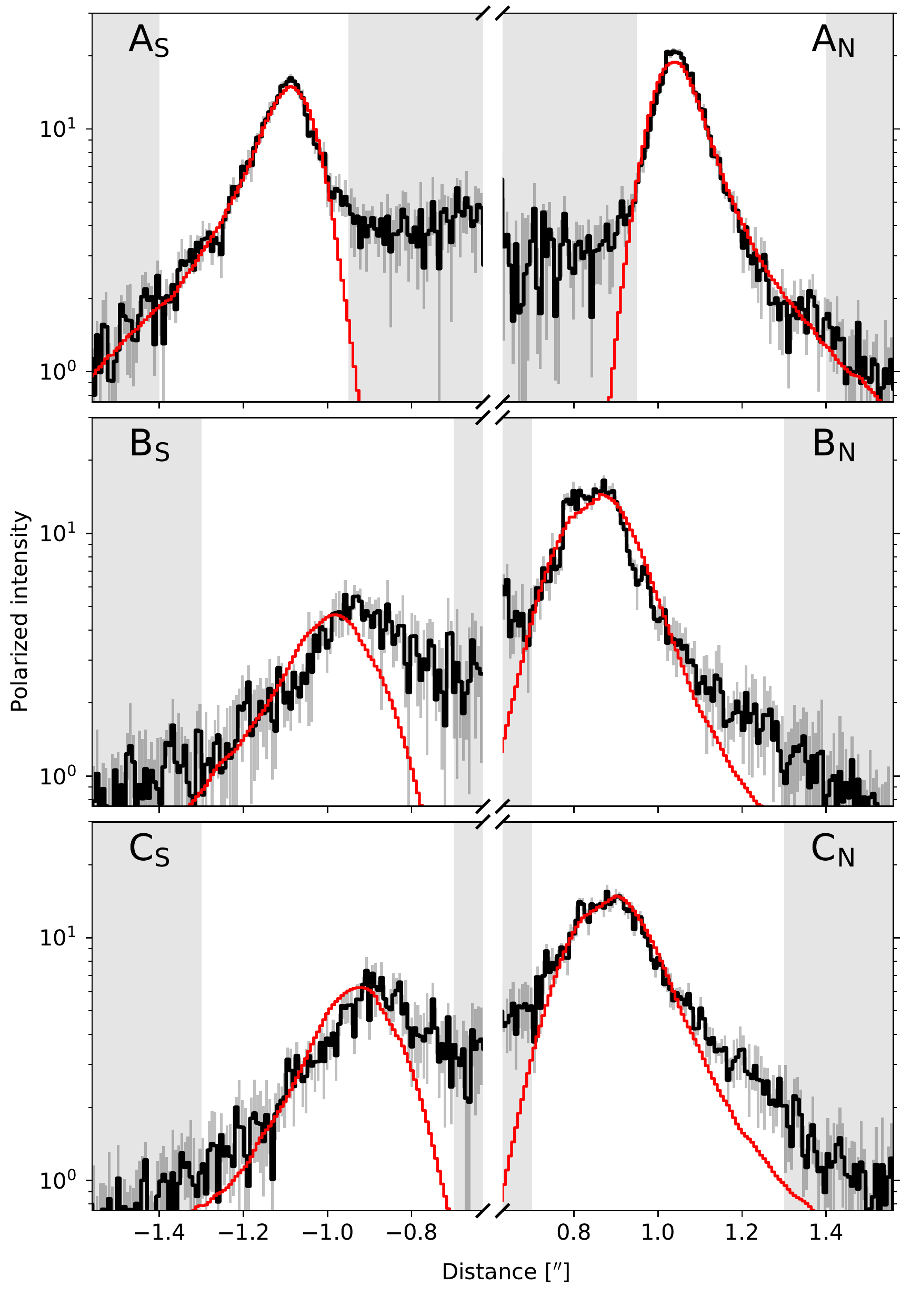}
    \caption{Radial profiles of the disk along the three cuts highlighted in Figure\,\ref{fig:cuts} (error bars are $1\sigma$), with the best-fit model over-plotted in red. The gray shaded areas show where the goodness of fit is \textit{not} estimated.}
\label{fig:results}
\end{figure}

\begin{figure*}
\centering
\includegraphics[width=\hsize]{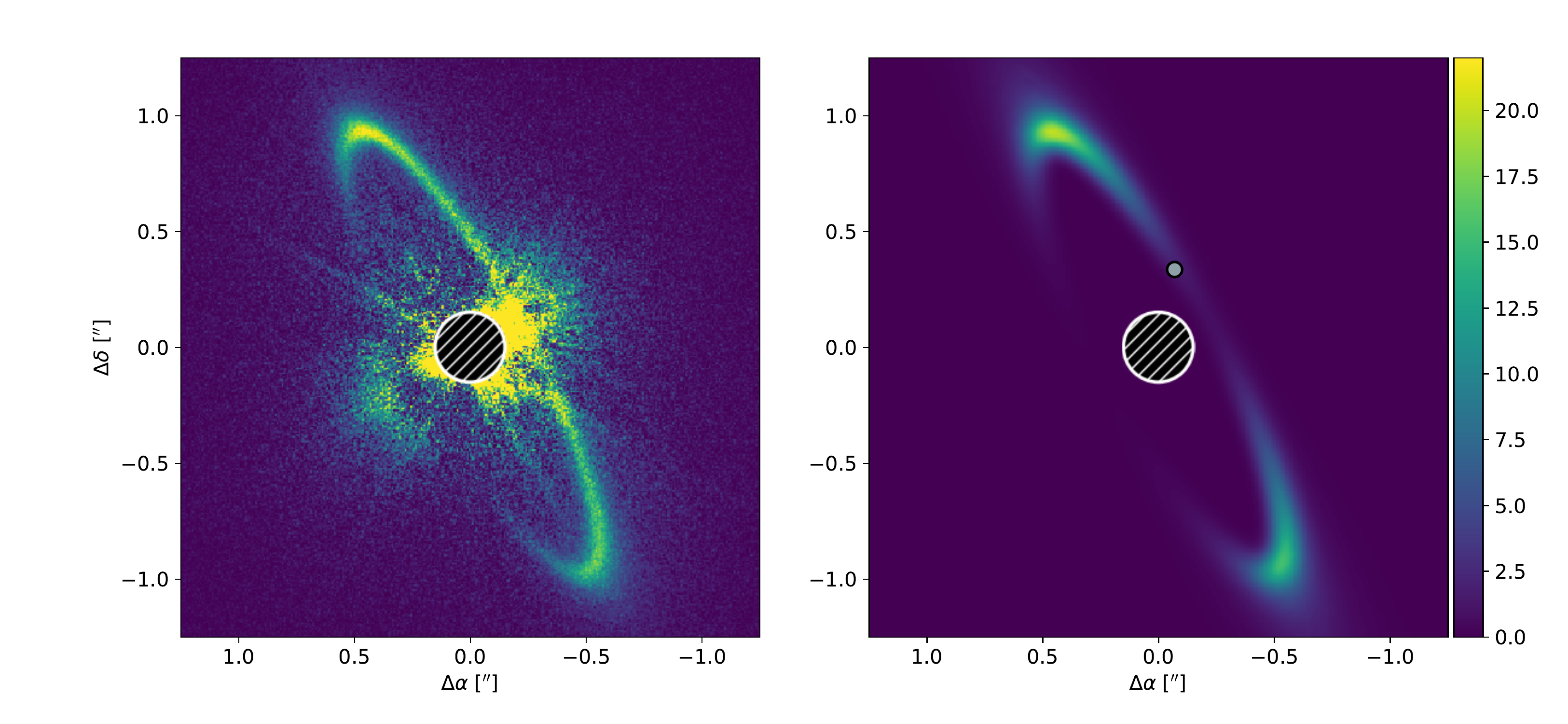}
\caption{Observations and best-fit model to the SPHERE/ZIMPOL observations of HR\,4796\,A, with the same linear stretch (left and right, respectively). For the model, the blue circle marks the location of the pericenter.}
\label{fig:images}
\end{figure*}

\begin{figure*}
\centering
\includegraphics[width=\hsize]{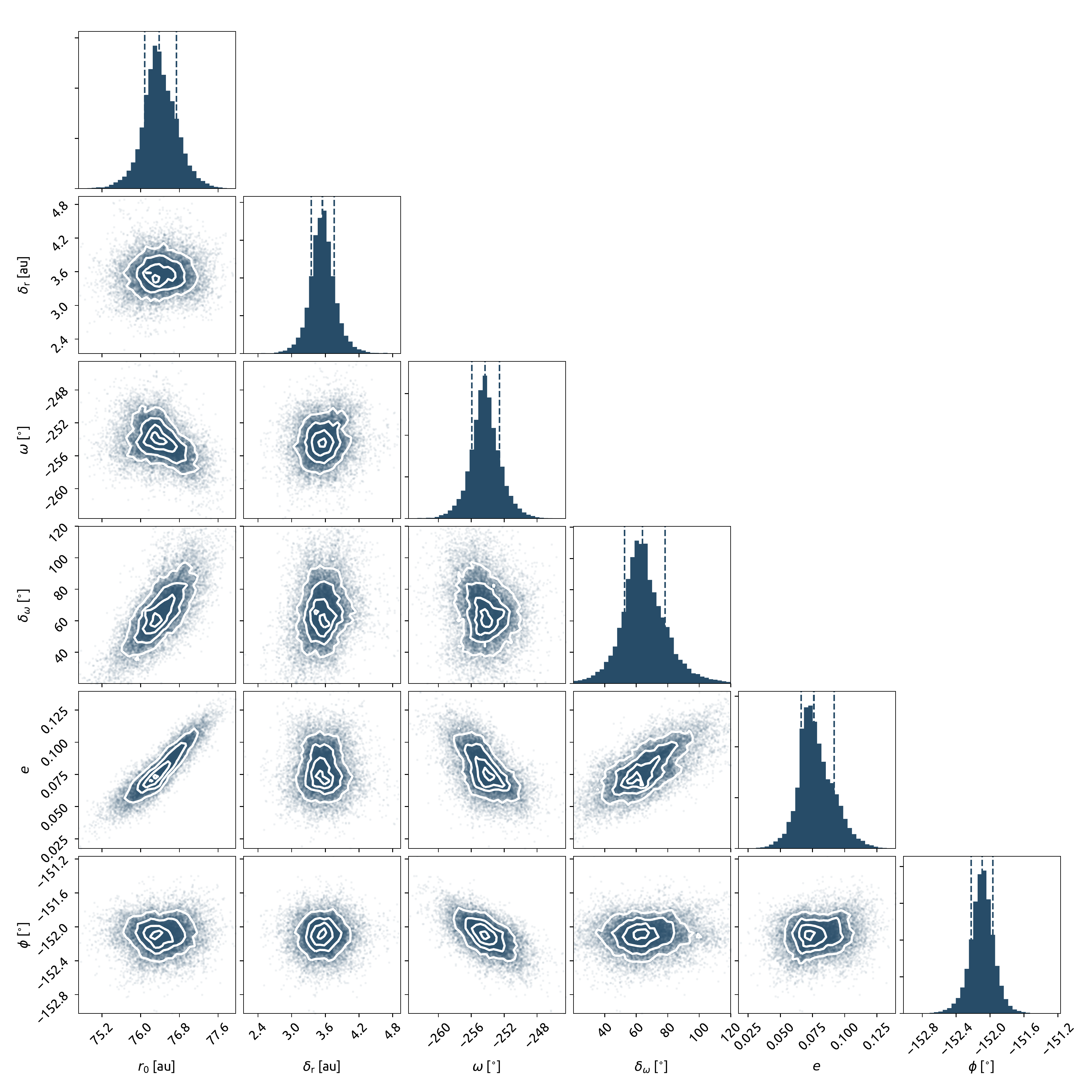}
    \caption{Projected probability density distributions along with the determined uncertainties for the different parameters in the modeling as well as density plots. The contours correspond to the $[0.12$, $0.39$, $0.68$, $0.86]$ density percentiles.}
\label{fig:corner}
\end{figure*}

The radial profiles are displayed in Fig.\,\ref{fig:results} and the observations and best-fit model are shown in Fig.\,\ref{fig:images}, with the same linear stretch (the difference in surface brightness closer to the semi-minor axis being due to the choice of the polarized phase function, see \citealp{Milli2019}). The location of the pericenter of the disk is marked by a blue circle on the right panel of Fig.\,\ref{fig:images}. While the best fit model slightly under-estimates the polarized intensity at separations larger than $1\arcsec$ along the B$_{\mathrm{N}}$ and C$_{\mathrm{N}}$ axis (still within $3\sigma$), and the peak positions along the B$_{\mathrm{S}}$ and C$_{\mathrm{S}}$ axis are not a perfect match to the data (even though the signal-to-noise is lower in those regions), overall, the radial profiles along the major axis (A$_{\mathrm{N}}$ and A$_{\mathrm{S}}$) are well reproduced, both for the peak positions and the slopes, up to $1\farcs4$. Figure\,\ref{fig:density} shows a top view of the weighted cross section of the best fit model.

The most probable parameters are summarized in Table\,\ref{tab:grid} and the probability density functions are shown in Figure\,\ref{fig:corner}. The uncertainties for the MCMC results are estimated from the $0.16$ and $0.84$ quartiles using the \texttt{corner} package (\citealp{ForemanMackey2016}). The projected posterior distributions are shown in Fig.\,\ref{fig:corner}. We find that the pericenter should be located on the front side of the disk, with $\omega = -254^{\circ}$$^{+2}_{-2}$. The reference radius of the disk is of $r_0 = 76.4^{+0.4}_{-0.3}$\,au, and the standard deviation of the radial distribution of the parent belt is $\dr = 3.6^{+0.2}_{-0.2}$\,au, while the standard deviation of the collisional distribution is $\dw = 63.9^{\circ}$$^{+14.4}_{-11.3}$. We find that the eccentricity is $e = 0.076^{+0.016}_{-0.010}$, and the position angle is $-152.1^{\circ}$$^{+0.1}_{-0.1}$, close to the value of $-151.6^{\circ}$ we previously found to define the location of the major axis.

\section{Discussion}\label{sec:discussion}

The debris disk around HR\,4796\,A has been resolved at high angular resolution on several occasions, with different instruments (e.g., \citealp{Lagrange2012,Rodigas2015,Milli2017,Schneider2018,Kennedy2018}). All these observations showed that the disk appears as a narrow ring. With the new ZIMPOL observations presented here, we also find the disk to be very narrow. To reproduce the observations, our modeling results suggest that the parent planetesimal belt follows a normal distribution with a standard deviation that can be as narrow as $3.6$\,au. One should note that the radial extent of the parent belt is of the order of the vertical of the disk, suggesting it is shaped as a thin torus. From the results of \citet{Rodigas2014}, \citet{Milli2017} concluded that the observed width of the disk around HR\,4796\,A could be explained by a planet lighter than Saturn, inwards of the ring, shepherding the debris disk (or even smaller if the planet is migrating, e.g., \citealp{Perez2019}). Given the comparable angular resolution provided by the SPHERE/IRDIS and ZIMPOL instruments, we do not revise the values reported in \citet{Milli2017} to explain the narrowness and eccentricity of the parent planetesimal belt. But the ZIMPOL observations provide new insights into the azimuthal distribution of the dust grains, as well as on the production of small dust grains in this young debris disk.

\subsection{The pericenter glow of HR 4796 A}

\begin{figure*}
\centering
\includegraphics[width=\hsize]{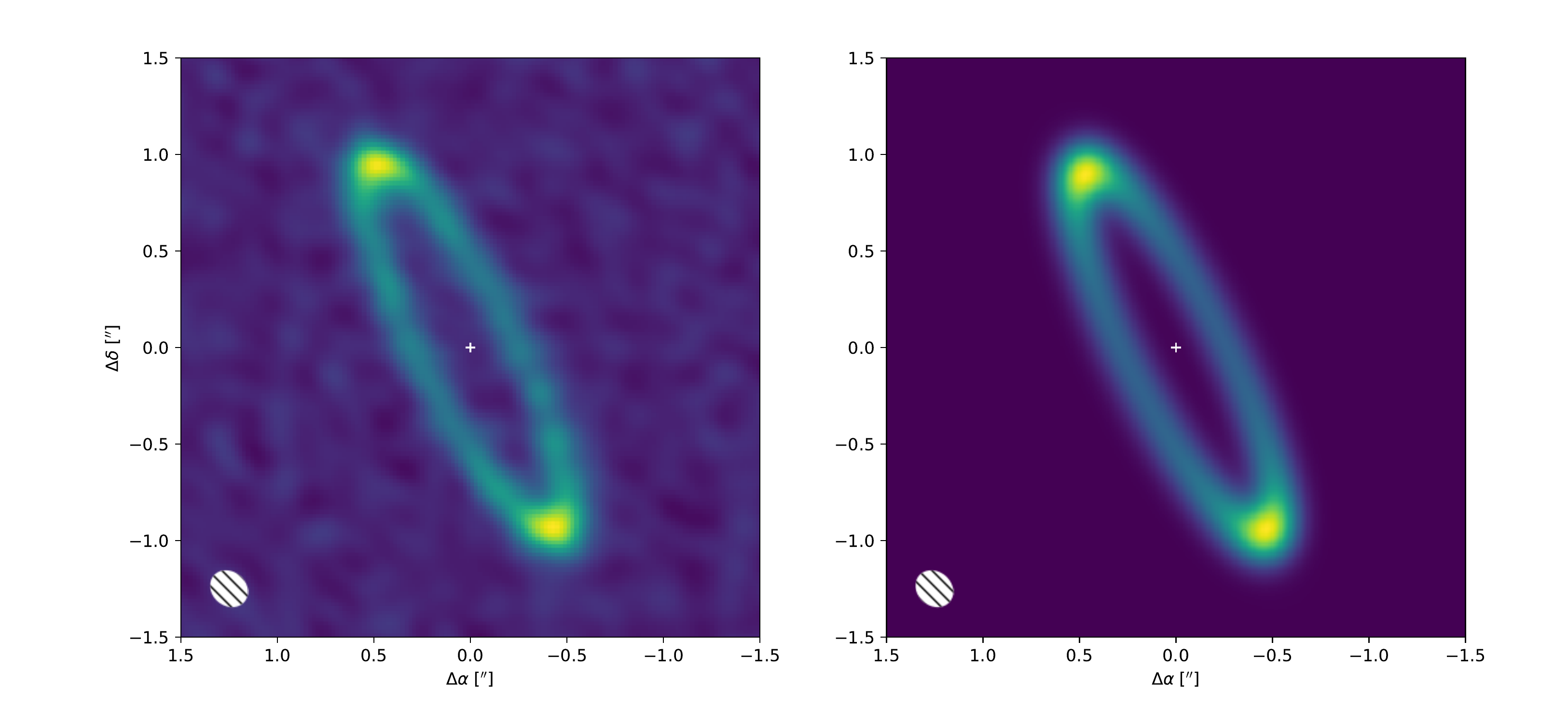}
\caption{\textit{Left:} ALMA $880$\,$\mu$m Briggs-weighted image of HR\,4796\,A. \textit{Right:} best fit model at the same wavelength, convolved with a similar beam (displayed in the bottom left corner of both images). No noise was added to the model.}
\label{fig:alma}
\end{figure*}

\begin{figure*}
\centering
\includegraphics[width=\hsize]{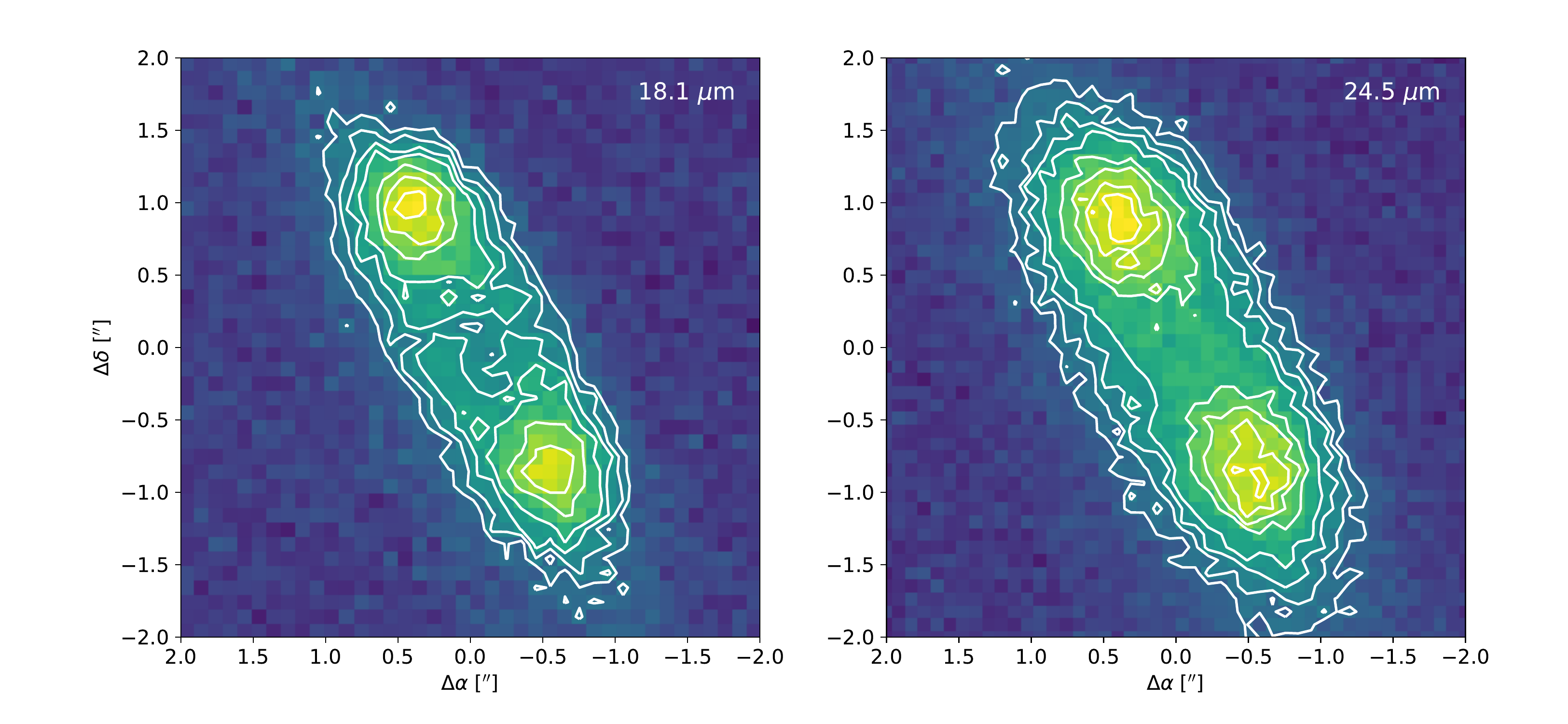}
\caption{Mock observations with the Michelle instrument ($18.1$\,$\mu$m, \textit{left} panel) and T-ReCS ($24.5$\,$\mu$m, \textit{right} panel).}
\label{fig:midir}
\end{figure*}

The pericenter glow effect was originally proposed by \citet{Wyatt1999} who modeled Keck\,II observations of HR\,4796\,A at $18.2\,\mu$m (also presented in \citealp{Telesco2000}). The asymmetry observed in thermal emission could be explained by the fact that the dust grains in the direction of the forced pericenter of the disk are closer to the star, and hence warmer as they receive more stellar light. The pericenter glow can also be observed in scattered light observations, the dust grains at the pericenter receiving, and therefore scattering, more light compared to the apocenter. \citet{Wyatt1999} reported an argument of periapsis of $26^{\circ}$ for an eccentricity of $0.02$, meaning that the pericenter is located close to the projected semi-major axis of the disk. Nonetheless, from their Fig.\,7 showing an ensemble of possible solutions, a forced eccentricity of $0.07$ would lead to an argument of periapsis of about $\sim75^{\circ}$ which, despite different reference system, is compatible with our results. However, \citet{Moerchen2011} modeled Michelle and T-ReCS mid-IR observations, and found that the pericenter should be located along the projected semi-major axis of the disk (perpendicular to the line of sight), but with an even larger eccentricity of $0.06$ compared to \citet[][$0.02$ when the pericenter is located near the projected semi-major axis]{Wyatt1999} . They also found a family of solutions, with increasing eccentricity when the pericenter is moved closer to the projected semi-\textit{minor} axis. However, since then it has been shown that this is not compatible with several studies that consistently found the pericenter to be located closer to the projected semi-\textit{minor} axis of the disk, without increasing $e$ too much (e.g., \citealp{Thalmann2011}; \citealp{Rodigas2014}; \citealp{Milli2017}, and this work). In Figure\,\ref{fig:dw0} we show the radial cuts along the semi-major axis of the disk, for a model in which particles are released uniformly in mean anomaly (all the other parameters being the same as the best model obtained before). One can note that in this model the brightness asymmetry is not well reproduced, justifying why our best fit model requires more particles to be released near the pericenter.

To further check the validity of our model, we computed mock observations at $18.1$ and $24.5$\,$\mu$m, to be compared with the observations presented in \citet{Moerchen2011}. These mock observations are shown in Figure\,\ref{fig:midir}; we used the same code as before, to compute images at the two wavelengths only considering thermal emission from the dust grains (using the same parameters as the best fit model to the optical observations), convolved them with 2D Gaussian with full width at half maximum of $0\farcs52$ and $0\farcs72$, and added noise to the images to mimic the real observations. While we did not aim at fitting those observations, visual inspection suggests that we are obtaining very comparable results, the brightness asymmetry being more pronounced at $18.1$\,$\mu$m than at $24.5$\,$\mu$m (where it is only marginally detected).

We additionally visually compared the ALMA observations presented in \citet{Kennedy2018} with a model computed using the same code at the same wavelength as the ALMA observations (thermal emission only). Figure\,\ref{fig:alma} shows the Briggs-weighted Band\,7 image (left) and the thermal emission for our best fit model, convolved by a 2D Gaussian similar to the beam of the observations. Our best fit model can reproduce the overall shape of the disk, the width of the ring, and the brightness distribution over all azimuthal angles. ALMA $880$\,$\mu$m observations being sensitive to larger grains, which are not subjected to strong radiation pressure, bound dust grains close to the cut-off size do not contribute significantly to the thermal emission at those wavelengths and the model traces the distribution of the parent planetesimal belt. For those large grains, even though they are preferentially released near the pericenter, their eccentricities are rather small (close to the eccentricity of the parent belt), and therefore when populating the orbits, dust particles are distributed almost uniformly in the azimuthal direction. The point being that, in our model, the larger the grains the less significant the value of $\dw$ becomes; ALMA observations can hardly constrain if dust grains are released in a preferential location of the disk (e.g., the pericenter in our case). Overall, this brings confidence to our best-fit model of the dust production and distribution in the disk around HR\,4796\,A, as it can reproduce (at least to the first order) observations from optical to millimeter wavelengths. In that model the pericenter glow effect plays a very minor role to explain the brightness asymmetry.

\subsection{Dust production around HR 4796 A}

The azimuthal number density of an eccentric ring should naturally peak at the location of the apocenter (\citealp{Pan2016}), and not at the location of the pericenter as is the case for HR\,4796\,A according to our best-fit model. As discussed in \citet{Pearce2014}, if a planet is interacting with a debris disk less massive than the planet, the number density should be higher at the apocenter. Such a model would therefore not be applicable to the disk around HR\,4796\,A (around which very stringent upper limits on the presence of companions have been placed, \citealp{Milli2017}). On the other side, \citet{Pearce2015} also investigated the case of the interactions between an eccentric planet and a debris disk of comparable mass. They conclude that the end-result of those interactions usually is a double-ringed debris disk, which is not observed for the disk around HR\,4796\,A.

A possible mechanism that could help explain our results would be the violent collision of large bodies at $76.4$\,au from the star (e.g., \citealp{Kral2015}). The numerical simulations presented in \citet{Jackson2014} indicate that all the bodies released from the original collision point will have to pass through the same location at each orbit, which would locally increase the collision probabilities. This collision point would therefore become the main production site for any secondary dust (or gas) in the system. \citet{Jackson2014} studied the cases of eccentric progenitors, and concluded that the resulting brightness asymmetry highly depends on where the collision took place. If the collision happened near the apocenter of the progenitors, then the brightness asymmetry is ``constructive'' due to the increased density at the collision point and the fact that particles spend more time near the apocenter (Fig.\,11-D of \citealp{Jackson2014}). On the other hand, if the collision took place close to the pericenter of the progenitors, then there is a competition\footnote{One should note however that the spatial resolution of the observations has to be taken into account here. If the disk is radially spatially resolved the flux is distributed over different areas (the apocenter being more extended), making the comparison less straightforward than for unresolved observations.} between the over-densities due to the ``pinch-point'' at the pericenter, and the more time spent by particles at the apocenter (Fig.\,11-C of \citealp{Jackson2014}). Nonetheless, the apocenter being more spread out than the pericenter (especially for the bound grains with the highest $\beta$ value), the latter one may still appear brighter than the former. Overall, the observations of the disk around HR\,4796\,A could be explained if the disk is the outcome of a unique, massive collision of initially eccentric progenitors (and the collision should have taken place close to the pericenter of the progenitors). Because the velocity dispersion is larger at the pericenter compared to the apocenter, a collision near the former could, in principle, generate a larger amount of small dust grains. Nonetheless, given how bright the disk is ($f_\mathrm{disk} = L_\mathrm{disk}/L_\star \sim 5 \times 10^{-3}$), the collision would have to have been an extremely rare event. If this is indeed the case, the implications for the formation of large planetesimals at distances larger than $70$\,au are strong (see \citealp{Kenyon2005}, as well as the discussion about the collisional status of the disk in \citealp{Kennedy2018}). One would need to form at least two very massive oligarchs in the outermost regions, on a timescale of a few Myr. Indeed, the mass of the body whose breakup is able to produce a disk of debris of fractional luminosity $5\times10^{-3}$ is extremely large. Taking the equations $2$-$5$ of \citet{Wyatt2007} linking $f_\mathrm{disk}$ to the mass of a collisional cascade producing it, we find that, even in the very optimistic hypothesis that all the mass is contained in $\leq1$m bodies, one needs at least a few Earth masses of material for the disk to be as bright as $f_\mathrm{disk} = 5\times10^{-3}$ at $75$\,au from its central star (\citealp{Augereau1999} had already found a similar estimate for the amount of $\leq1$m bodies). As a consequence, the catastrophic breakup scenario requires the breakup of planetary object, probably at least in the super-Earth range. One may furthermore wonder if such a collision should not also have released significant amount of gas (as postulated for $\beta$\,Pictoris for instance, \citealp{Dent2014}). Despite sensitive observations, no gas has been detected in the outermost regions of HR\,4796\,A (\citealp{Kennedy2018}), but \citet{Iglesias2018} reported possible ``falling evaporating bodies'', by detecting variable extra absorption lines in optical spectroscopic observations.

\subsection{The effect of HR 4796 B on the debris disk}\label{sec:HR4796B}

\begin{table}
\caption{Stellar properties for HR\,4796\,A and HR\,4796\,B.}
\label{tab:gaia}
\centering
\begin{tabular}{lcc}
\hline\hline
Parameters & HR\,4796\,A & HR\,4796\,B \\
\hline
$\alpha$                        & $189.0039\pm0.0977$ & $189.0019\pm0.0472$ \\
$\delta$                        & $-39.8696\pm0.1004$ & $-39.8711\pm0.0518$ \\
$\pi$ [mas]                     & $13.9064\pm0.1349$  & $14.1030\pm0.0625$ \\
$\mu_{\alpha}$ [mas.yr$^{-1}$]  & $-55.653\pm0.181$   & $-59.236\pm0.096$ \\
$\mu_{\delta}$ [mas.yr$^{-1}$]  & $-23.740\pm0.230$   & $-29.867\pm0.125$ \\
R$_{\mathrm{V}}$ [km.s$^{-1}$]  & $7.10\pm1.10$       & $7.63\pm0.70$  \\
\hline
\end{tabular}
\end{table}

To estimate if the M-type star HR\,4796\,B can reasonably have an effect on the disk around HR\,4796\,A (projected separation of $568.3$\,au), through radiation pressure, stellar winds, or gravitational interactions (e.g., \citealp{Thebault2010}, \citealp{Cuello2019}), we first check the separation between both stars. We used their Gaia DR2 measurements (\citealp{Gaia2018}), which are reported in Table\,\ref{tab:gaia}. The radial velocity of HR\,4796\,A was taken from \citet{Iglesias2018}, and we estimated the one of HR\,4796\,B from UVES observations (program IDs 082.C-0218 and 089.C-0207). The first noteworthy difference is the parallaxes of both stars, which translate to distances of $71.9$ and $70.9$\,pc for HR\,4796\,A and B, respectively. However, the measurements for HR\,4796\,A have large uncertainties due to its brightness and the star is flagged for possible astrometric errors. Therefore, we here assume that the B star has the same parallax as the A star and to evaluate if the former can have an impact on the disk around the A star, we checked if the system is bound. To that end, we estimated the escape velocity of B with respect to A, as $\sqrt{2 G M_{\star, \mathrm{A}} / r}$, where $G$ is the gravitational constant, $M_{\star, \mathrm{A}}$ is the mass of the HR\,4796\,A ($1.3$\,M$_{\odot}$), and $r$ the separation between the two stars. With the positions and velocities of both stars, we find an escape velocity of $2.01$\,km.s$^{-1}$, compared to a relative velocity of $2.47$\,km.s$^{-1}$ (estimated from the proper motions and radial velocities of both stars). As we assumed the same parallax for both stars, this is the most favorable case with the smallest three-dimensional separation. Therefore, with the available astrometric measurements, it seems that HR\,4796\,A and B are not bound to each other, most likely minimizing the possible impact that the B component can have on the debris disk, but this may have to be revisited with a more reliable astrometry for HR\,4796\,A in the near future.

\subsection{An asymmetric halo of small dust grains}

\citet{Schneider2018} presented deep HST observations of the disk around HR\,4796 and revealed an extremely extended halo outside of the birth ring. They detect this ``exo-ring'' material up to $\sim 10\arcsec$ along the north-east side, while the south-west side appears more compact. As a matter of fact, there seems to be an ``arc-like'' feature along the south-west side, which could be due to either interactions with the ISM gas, or with HR\,4796\,B. Based on the discussion above, the latter scenario is probably unlikely, if HR\,4796\,A and B are indeed not gravitationally bound. The authors mention the possibility that the dust grains in the halo \textit{may} be unbound from the central star, but those dust grains should be evacuated from the system extremely quickly (see e.g., \citealp{Thebault2008}), and therefore should hardly be detected. On the other hand, bound grains, with large $\beta$ ratios can have their apocenter at $10-20$ times the separation of the birth ring. Therefore, in the following, we will attempt to reproduce the general shape of the exo-ring (extended emission along the north-east with an arc-like shape along the south-west), based on our current best-fit model, and we will be considering bound dust grains with high $\beta$ ratios. This subsection is meant to be speculative, given the complexity of the problem and we simply aim at providing two scenarios to explain the HST observations. 

First of all, out-of-the-box, our current best fit model cannot reproduce the HST observations. We find that the pericenter is located along the north side, and therefore, the disk will naturally be more extended in the direction toward the apocenter, i.e., along the south side, while the HST observations show otherwise. We will therefore consider here two alternative explanations: \textit{(i)} the dust grains set on high eccentricity orbits are interacting with the local interstellar medium, and \textit{(ii)} a slow precession of the pericenter of the debris disk is causing the apparent asymmetry.

\subsubsection{Description of the code}

When considering additional forces, such as gas drag, the analytical prescriptions given in Eq.\,\ref{eqn:orbit} are no longer useful to estimate the orbital parameters of the dust grains. Therefore, we opted to use a simple Runge–Kutta integrator at the 4th order to estimate the positions and velocities of the particles. The forces considered here are the gravitational potential from the central star (weighted by $1-\beta$ when considering the effect of radiation pressure), and the gas drag on each dust grain. We followed the work of \citet{Marzari2011} (see also \citealp{Pastor2017}) for the implementation, and the acceleration felt by a dust grain is estimated as
\begin{equation}\label{eqn:ism}
\boldsymbol{f} = -C_{\mathrm{D}} n_{\mathrm{H}} m_{\mathrm{H}} A_{\mathrm{s}} \lvert \boldsymbol{v} - \boldsymbol{v_{\mathrm{H}}}\rvert (\boldsymbol{v} - \boldsymbol{v_{\mathrm{H}}}),
\end{equation}
where $\boldsymbol{v}$ is the velocity of the dust grain, $\boldsymbol{v_{\mathrm{H}}}$ the velocity of the ISM gas, $n_{\mathrm{H}}$ and $m_{\mathrm{H}}$ are the density and mass of the hydrogen atoms in the ISM, and $C_{\mathrm{D}}$ is a drag coefficient (set to $2.5$ as in \citealp{Marzari2011}). The cross section of a dust grain is given by $A_{\mathrm{s}} = \pi s^2$.

To produce an image, we initially release $10\,000$ particles (uniformly distributed in mean anomaly) within the birth ring (described by its semi-major axis, eccentricity, and argument of pericenter, which are the same as the best fit model) and follow their trajectories in time. For the sake of simplicity, in those simulations there is no preferential collisions near the pericenter. At the initialization of the simulation, we first check if each particle is indeed bound to the system, by estimating its initial velocity, and comparing it to the escape velocity of the system. Only particles that are bound are kept in the simulation. At each time step (usually $4$\,years) we save the $(x,\,y,\,z)$ positions of each of the particles, and then project those values onto the sky plane, depending on the inclination, position angle, and opening angle of the disk. For all the simulations presented in the following, we assume the Henyey-Greenstein approximation for the phase function, with $g = 0$ (isotropic scattering), and compute the surface brightness by multiplying the phase function by $Q_\mathrm{sca} \pi s^2 / (4\pi r^2)$. Also, we only consider a single grain size of $s = 7.15$\,$\mu$m, as this is the smallest size for which all the particles remain bound to the star, and have a high $\beta$ ratio of $\sim 0.46$.

\subsubsection{Interaction with local ISM gas}

\begin{figure*}
\centering
\includegraphics[width=\hsize]{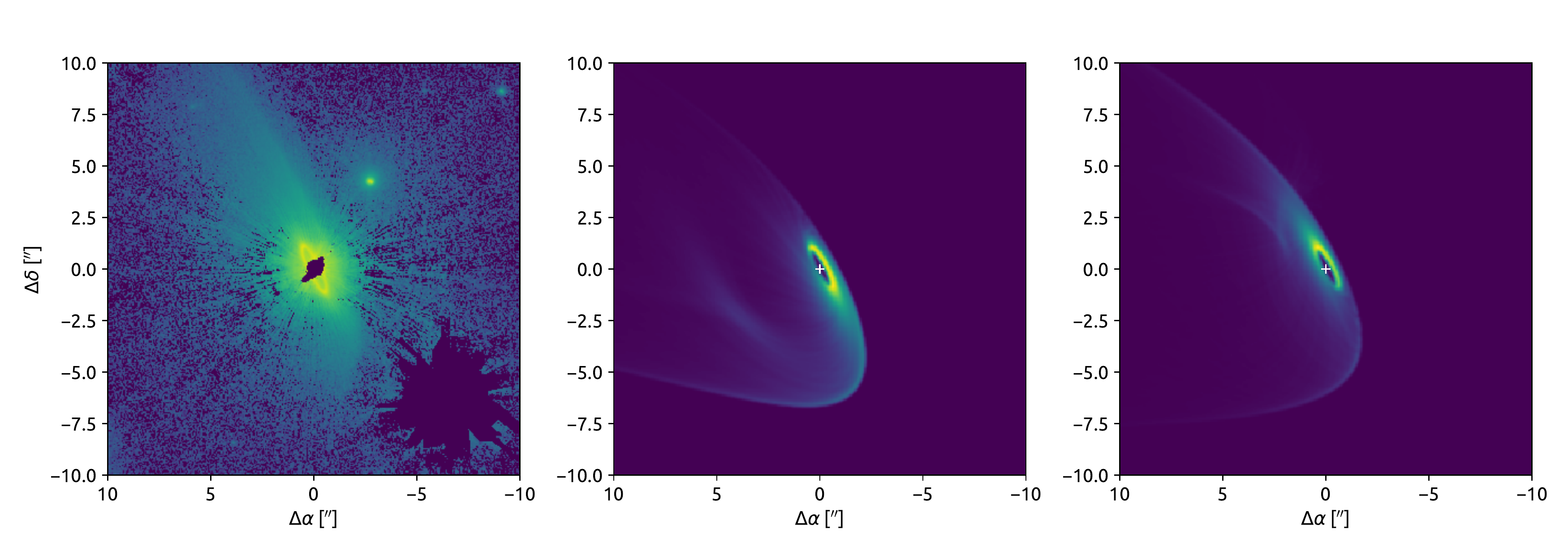}
    \caption{HST observations, as published in \citet{Schneider2018} (\textit{left} panel), and mock HST observations calculated from the N-body simulations of small, bound, dust grains around HR\,4796\,A, when considering interactions with the local ISM (clockwise and counter-clockwise rotations on the \textit{middle} and \textit{right} panels, respectively). The images are in log-scale.}
\label{fig:ism}
\end{figure*}

The simulations presented in this section last for $100\,000$\,years ($25\,000$ steps of $4$\,years), and the free parameters that we investigate are: the density of hydrogen atoms in the ISM $n_{\mathrm{H}}$, the direction and velocity of the ISM gas (all encompassed in the $\boldsymbol{v_{\mathrm{H}}}$ vector), and the direction of rotation of the dust grains in the disk. To simplify the definition of the problem, we assume that the $\boldsymbol{v}$ vector only represents the orbital velocity of the particles, while the $\boldsymbol{v_{\mathrm{H}}}$ vector contains information about both the proper motion of the star and the direction of motion of the ISM gas. As mentioned before, a full exploration of the parameter space is out of the scope of this paper, and we simply aim at providing a qualitative assessment on how to explain the HST observations.

Figure\,\ref{fig:ism} shows two simulations as well as the HST observations published in \citet{Schneider2018} (left panel), where the only difference is the rotation direction of the dust grains in the disk (clockwise and counter-clockwise for the middle and right panels, respectively). The images have been convolved by a 2D gaussian with a standard deviation of $1$\,pixel, and are shown with a log stretch to highlight the faint outer regions (as a consequence, the birth ring appears broader than in ZIMPOL simulations which are shown with a linear scale). For those simulations, we tried to fix as many free parameters as possible. We therefore assumed that the radial velocity of the ISM gas matches the one of the central star (\citealp{Iglesias2018} detected several absorption lines at different velocities, but we cannot know the relative distances of those clouds, only that they are between the star and the observer). Furthermore, given that the arc-like shape is in the east-west direction, we assumed that the ISM gas has a null velocity in the north-south direction, and that the $\delta$ component of the $\boldsymbol{v_{\mathrm{H}}}$ vector is equal to the proper motion of the star (see Table\,\ref{tab:gaia}). This leaves us with $\alpha$, $n_{\mathrm{H}}$, and the direction of rotation of the dust grains in the disk as free parameters.

Similarly to \citet[][studying the disk around HD\,61005]{Pastor2017}, we find that the ISM density has to be quite significant to produce the arc-like feature in the south-west direction (but overall, the density and the amplitude of the velocity are degenerate parameters). For the simulations presented in Figure\,\ref{fig:ism}, we set $n_{\mathrm{H}} = 125$\,cm$^{-3}$ and $\alpha = -100$\,mas.yr$^{-1}$. This means that in those simulations, the star is moving $55$\,mas.yr$^{-1}$ towards the west (following its proper motion in right ascension) and that the ISM gas is moving $45$\,mas.yr$^{-1}$ towards the east. Overall, both simulations can produce the arc-shape structure as seen with HST, roughly at the same separation ($\sim 5\arcsec$), but when the dust grains are moving in the clockwise direction onto the sky plane, the north-east side of the disk is much fainter, and the south side of the birth ring appears brighter than the north side. This suggests that \textit{if} the extended halo is indeed the consequence of interactions with the ISM gas, then, the dust grains would most likely be orbiting counter-clockwise around HR\,4796\,A. The main caveat in this scenario is the rather large density of the ISM gas required, compared to the surroundings of the solar system,, similarly to what was found for HD\,61005 by \citet{Pastor2017}. However, the volume density of the cold, dense, interstellar medium can vary between $5-100$\,cm$^{-3}$ (e.g., \citealp{Nguyen2019} and references therein). Therefore, our results, while on the higher end of the range, remain overall compatible with studies of the interstellar medium.

\subsubsection{Precession of the pericenter}

\begin{figure}
\centering
\includegraphics[width=\hsize]{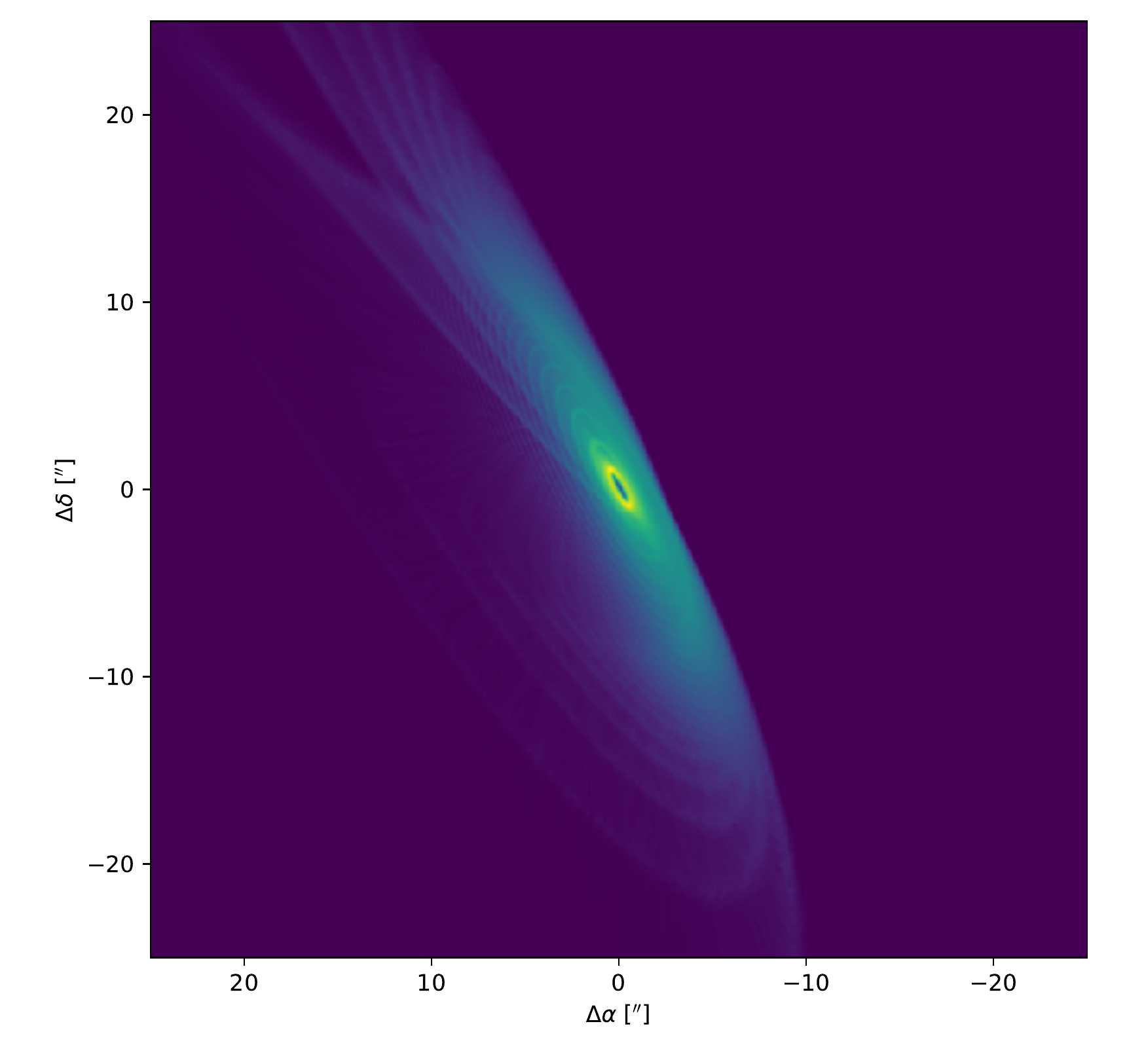}
\caption{Mock HST observations calculated from the N-body simulations of small, bound, dust grains around HR\,4796\,A, when considering that the pericenter is precessing over time. The image is in log-scale.}
\label{fig:precession}
\end{figure}

The second scenario we investigate to explain the HST observations is based on the fact that an eccentric disk is more radially extended towards the apocenter. Several studies of the disk around HR\,4796\,A consistently found that the pericenter of the disk is located close to the (projected) semi-minor axis of the disk, on the north side. But, if the disk is precessing (see also \citealp{Lohne2017}), in the past the pericenter might have been located on the south side, resulting in a more extended disk along the north side. To test this hypothesis, we run similar N-body simulations, for a single grain size, without any additional forces (only gravitational and radiation pressure forces). The two free parameters of the simulations are the total duration $t = t_\mathrm{sim}$ and the precessing rate $\Omega$ of the pericenter (i.e., $(\omega_{\mathrm{end}} - \omega_{\mathrm{start}}) / t_\mathrm{sim}$). In this case, we assume that the pericenter of the disk is rotating counter-clockwise, from the south-west towards the north-west (just past the semi-minor axis of the disk). Therefore, we only consider dust grains that are also rotating counter-clockwise around the star. 

We chose $\omega_{\mathrm{start}} = -320^{\circ}$, $\omega_{\mathrm{end}} = -254^{\circ}$, and a total duration $t_\mathrm{sim} = 200\,000$\,years, divided in $n_\mathrm{sim} = 500$ different steps. At $t = 0$, the pericenter is located at $\omega_{\mathrm{start}}$, and the N-body simulation last over $t_\mathrm{sim}$, with a time step of $4$\,years and we save the position of all the particles for the last $20$ iterations (hence the last $80$\,years). The $i^{\mathrm{th}}$ N-body simulation (out of the $500$) will last slightly less time ($t_\mathrm{sim} - i \times t_\mathrm{sim}/n_\mathrm{sim}$) and the location of the pericenter will have moved slightly ($\omega_{\mathrm{start}} + i \times (\omega_{\mathrm{end}} - \omega_{\mathrm{start}})/n_\mathrm{sim}$), and we still save the last $80$\,years of the simulation. At the end, we simply collapse all the $n_\mathrm{sim}$ images together (see \citealp{Thebault2012b} for a similar approach). 

Figure\,\ref{fig:precession} shows the result of this simulation (on an aggressive log stretch to reveal the outermost regions, also rendering the birth ring thicker than in other Figures). One can see that indeed the south-west side is slightly dimmer than the north-east side, less extended in the radial direction, and that arc-like structures start to develop in the south-west side. The concentric ellipses close to the birth ring are due to the fact that we save the positions of the particles over the last $80$\,years of the simulations, but the time step between each N-body simulations is larger than $80$\,years ($400$\,years). It also appears, that we cannot reproduce the scale of the HST image in this simulation (the arc shape is located at about $5\arcsec$ in the HST observations, while the disk still appears bright at even $10\arcsec$). This would most likely suggest that we are observing grains with a $\beta$ value slightly smaller than $0.46$.

This approach is extremely simplistic, and heavily depends on the initial conditions and the duration of the simulation (i.e, when we decide to stop the simulation). The main caveat is that we assume that the dust particles can actually survive over $t_\mathrm{sim}$ without being destroyed. If we assume $\beta = 0.45$, $e = 0.07$, $a = 76.4$\,au, and draw uniformly $1\,000$ values for the true anomaly between $[0, 2\pi)$, we can estimate the orbital parameters of the dust grains following Eq.\,\ref{eqn:orbit}. We can then estimate the orbital period of those grains as $T = 2\pi \sqrt{a_{\mathrm{n}}^3/GM_\star(1 - \beta)}$. We find a mean value of $18\,500$\,years with a standard deviation of $15\,800$\,years (the distribution is strongly peaked at $\sim 5\,500$\,years). This means, that over $t_\mathrm{sim} = 200\,000$\,years, the dust grains would, on average, pass trough the birth ring $10-40$ times (this criterion, combined with the CPU time for the simulation, drove the choices for the precession rate and the total duration of this simple exercise). Depending on the optical depth of the birth ring, those grains may get destroyed before that, which may work in our favor. Indeed, to produce an arc-like feature at large distances from the star, one needs to break the symmetry. If the disk has been precessing for a long time, and none of the particles are destroyed, then the result would most likely be symmetric. However, if after several orbits, the oldest dust grains are destroyed in the birth ring, this can generate a possible asymmetry in the surface brightness of the disk, as we ``lose memory'' from past events.

Overall, our simulation does not properly match the observations, and remains speculative, but it seems to be going in the right direction to try and explain the HST observations presented in \citet{Schneider2018}. Those observed images could well be result of a combination of both effects, precession of the disk and interactions with local interstellar medium. Furthermore, the scenario that we invoked earlier to explain the morphology of the disk (a collision at $76.4$\,au, following the work of \citealp{Jackson2014}), is not incompatible with precession of the disk. \citet{Jackson2014} mention that even though the original collision point remains static on orbital time-scales, it should still be able to precess, due for instance to the presence of other massive bodies in the vicinity of the birth ring. Further investigation of the lifetime of the dust grains, precession rate, and the reason for the precession of the disk is out of the scope of this paper, the intent here being simply to propose a scenario that can reconcile all the available observations of the disk.

\section{Conclusion}

We presented high angular resolution SPHERE/ZIMPOL observations that reveal the bright debris disk around HR\,4796\,A with exquisite angular resolution (see also \citealp{Milli2019} for a discussion about the polarized phase function). At optical wavelengths, the north-east side of the disk is brighter than the south-west side, which we aimed at reproducing in this paper. We modeled the radial profiles along (and close to) the semi-major axis of the disk with a code that includes a simple prescription of the effect of radiation pressure. With this code, which is faster (but also much simpler) than models including a more accurate treatment of collisional and dynamical evolution of debris disks (e.g., \citealp{Kral2013,Kral2015,Lohne2017}), we are able to reproduce the observed profiles on both sides of the disk. We can reproduce the outer edge of the disk without invoking the presence of an outward planet truncating the disk (similarly to \citealp{Thebault2012}) and we find that the underlying planetesimal belt can be as narrow as a few au. As previously stated in the literature, this could be related to the presence of an unseen planet, inwards of the planetesimal belt, shepherding the debris disk. We find, similarly to other studies in the past years, that the pericenter of the disk is located close to the projected semi-minor axis of the disk. We show that with such a configuration, the pericenter glow, that had been postulated to explain marginally resolved observations, has in fact very little impact on the azimuthal brightness distribution of the disk. To reproduce the observed brightness asymmetry, we find that small dust grains must be preferentially released, as a result of collisions between larger bodies, close to the pericenter of the disk. Finally, our best-fit model can self-consistently reproduce most of the available observations of the disk, from optical to millimeter wavelengths. The only dataset that remains challenging to explain are the recently published HST observations that reveal an extended halo at large separations from the star. After concluding that HR\,4796\,B \textit{may} not be bound with HR\,4796\,A, thus minimizing its possible effect on the disk, we propose two possible scenarios that could help explaining the HST dataset; interactions with the local interstellar medium and precession of the disk. Even though the results remain speculative, those hypotheses (or a combination of those) could help reproduce the very faint halo of small dust grains. But further investigation is required to confirm our findings, for instance by performing more refined dynamical simulations (e.g., accounting for grain-grain collisions during the evolution of an eccentric disk, \citealp{Lohne2017})

\begin{acknowledgements}
We thank Glenn Schneider for kindly providing the HST observations. We thank the anonymous referee for providing useful comments that helped improving the paper, especially the description of the code.
This research has made use of the SIMBAD database (operated at CDS, Strasbourg, France) and of the Spanish Virtual Observatory (http://svo.cab.inta-csic.es, supported from the Spanish MICINN / MINECO through grants AyA2008-02156, AyA2011-24052). This research made use of Astropy, a community-developed core Python package for Astronomy (\citealp{Astropy}), as well as the TOPCAT software (\citealp{Taylor2005}).
SPHERE is an instrument designed and built by a consortium consisting of IPAG (Grenoble, France), MPIA (Heidelberg, Germany), LAM (Marseille, France), LESIA (Paris, France), Laboratoire Lagrange (Nice, France), INAF–Osservatorio di Padova (Italy), Observatoire de Gen\`eve (Switzerland), ETH Zurich (Switzerland), NOVA (Netherlands), ONERA (France) and ASTRON (Netherlands) in collaboration with ESO. SPHERE was funded by ESO, with additional contributions from CNRS (France), MPIA (Germany), INAF (Italy), FINES (Switzerland) and NOVA (Netherlands).  SPHERE also received funding from the European Commission Sixth and Seventh Framework Programmes as part of the Optical Infrared Coordination Network for Astronomy (OPTICON) under grant number RII3-Ct-2004-001566 for FP6 (2004–2008), grant number 226604 for FP7 (2009–2012) and grant number 312430 for FP7 (2013–2016). We also acknowledge financial support from the Programme National de Plan\'etologie (PNP) and the Programme National de Physique Stellaire (PNPS) of CNRS-INSU in France. This work has also been supported by a grant from the French Labex OSUG@2020 (Investissements d'avenir – ANR10 LABX56). The project is supported by CNRS, by the Agence Nationale de la Recherche (ANR-14-CE33-0018). It has also been carried out within the frame of the National Centre for Competence in Research PlanetS supported by the Swiss National Science Foundation (SNSF). MRM, HMS, and SD are pleased to acknowledge this financial support of the SNSF. Finally, this work has made use of the SPHERE Data Centre, jointly operated by OSUG/IPAG (Grenoble), PYTHEAS/LAM/CESAM (Marseille), OCA/Lagrange (Nice) and Observatoire de Paris/LESIA (Paris). We thank P. Delorme and E. Lagadec (SPHERE Data Centre) for their efficient help during the data reduction process. 
J.~O., A.~B., J.\,C.~B., D.~I., M.~M., M.\,R.~S., and C.~Z. acknowledge support from the ICM (Iniciativa Cient\'ifica Milenio) via the Nucleo Milenio de Formación planetaria grant. J.~O. acknowledges support from the Universidad de Valpara\'iso and from Fondecyt (grant 1180395). A.~B. acknowledges support from Fondecyt (grant 1190748). 
M.~M. acknowledges financial support from the Chinese Academy of Sciences (CAS) through a CAS-CONICYT Postdoctoral Fellowship administered by the CAS South America Center for Astronomy (CASSACA) in Santiago, Chile.
FMe acknowledges funding from ANR of France under contract number ANR-16-CE31-0013.
J.\,C.~B. acknowledges support from Proyecto FONDECYT postdoctorado 2018 nro. 3180716.
G.~M.~K. is supported by the Royal Society as a Royal Society University Research Fellow. 
A.~Z. acknowledges support from the CONICYT + PAI/ Convocatoria nacional subvenci\'on a la instalaci\'on en la academia, convocatoria 2017 + Folio PAI77170087. 

\end{acknowledgements}

\bibliographystyle{aa}

\appendix

\section{Miscellaneous}

\begin{figure}
\centering
\includegraphics[width=\hsize]{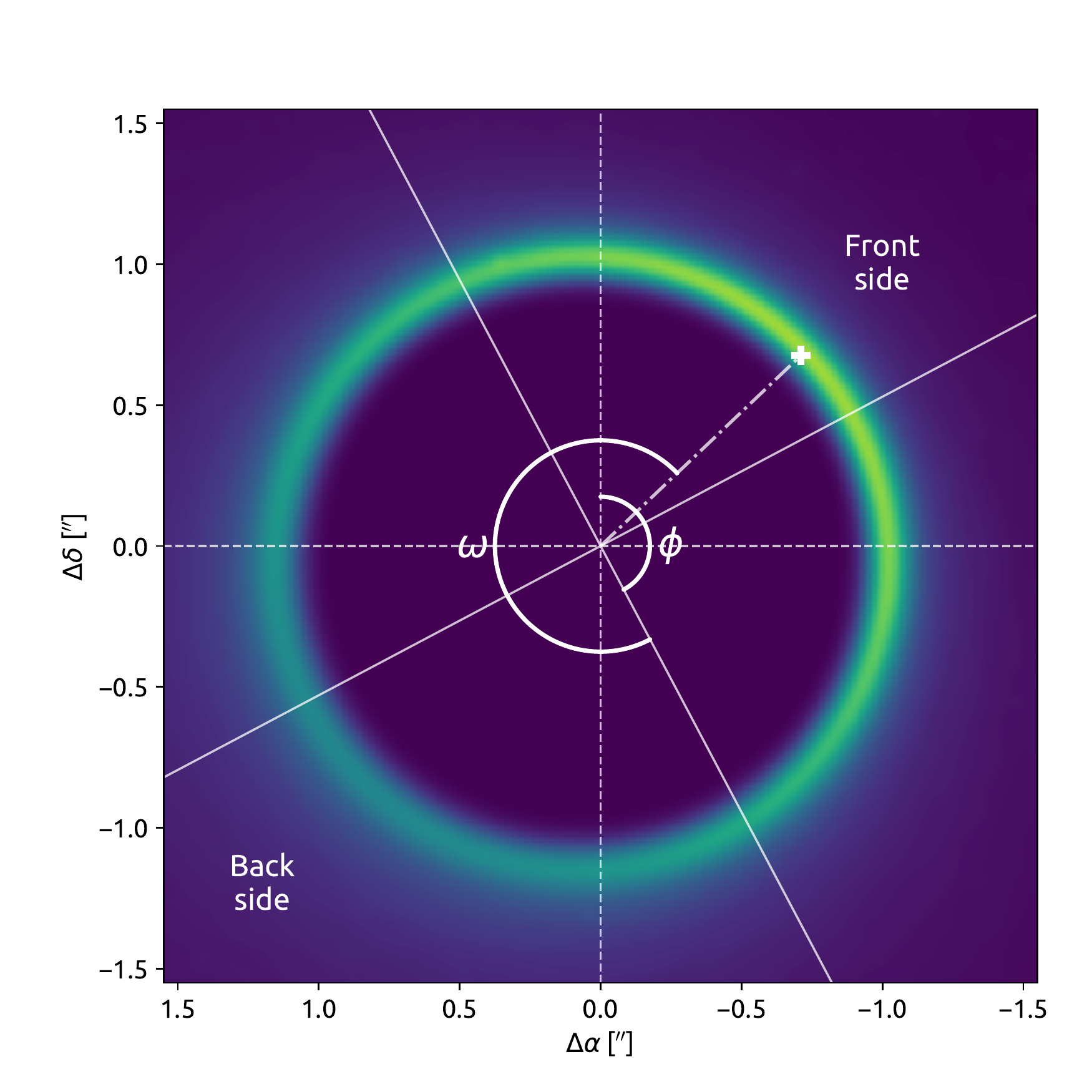}
\caption{Top view of the weighted cross section of the best fit model, indicating the position angle and argument of periapsis.}
\label{fig:density}
\end{figure}

Figure\,\ref{fig:density} shows the de-projected, weighted cross section, of the best fit model to the ZIMPOL observations, also displaying how we define the argument of periapsis and the position angle of the disk.

\begin{figure}
\centering
\includegraphics[width=\hsize]{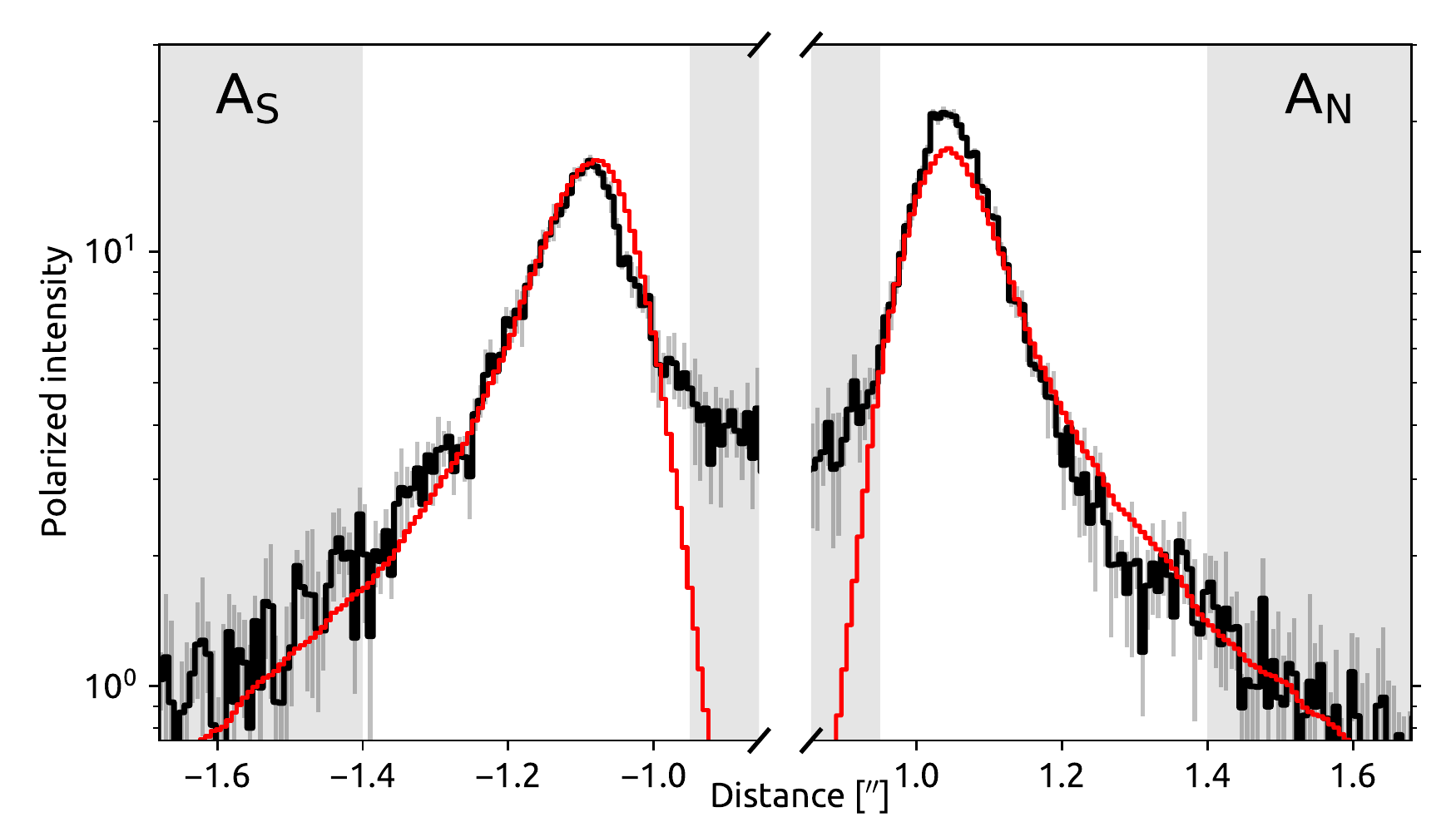}
\caption{Cut along the semi-major axis of the disk when particles are released uniformly in mean anomaly in the disk (i.e., no preferential collision at the pericenter).}
\label{fig:dw0}
\end{figure}

Figure\,\ref{fig:dw0} shows the cut along the semi-major axis (profiles A$_\mathrm{N}$ and A$_\mathrm{S}$) for a model in which the particles are not released from a preferential location in the disk. All the parameters of the model are the same as the best fit model to the ZIMPOL observations otherwise. This demonstrates that for our best fit solution, pericenter glow alone cannot fully reproduce the brightness asymmetry observed along the major axis of the disk. The other radial cuts are not shown as the main point of the Figure is to discuss the contribution of pericenter glow to the brightness distribution.

\end{document}